\documentclass[journal]{IEEEtran}

\usepackage[table]{xcolor}
\usepackage{cite}
\usepackage{graphicx}
\usepackage{lipsum}
\usepackage{stfloats}
\usepackage{bm}
\usepackage{amsmath,amssymb}
\usepackage{amsthm}
\usepackage{pifont}
\usepackage{bbding}
\usepackage{booktabs}
\usepackage{ctable}
\usepackage{threeparttable}
\usepackage{footmisc}
\usepackage{framed} 
\usepackage{array}
\usepackage{multirow}
\usepackage{longtable}
\usepackage{rotating}
\usepackage{fancyhdr}
\usepackage{lastpage}
\usepackage{layout}
\usepackage{wrapfig}
\usepackage{xcolor}
\usepackage{tabularx}
\usepackage{tabu}
\usepackage{enumerate}
\usepackage{enumitem}
\usepackage{url}
\usepackage{dsfont}
\usepackage{algorithm}
\usepackage{algorithmic}
 \usepackage{enumerate}
 \usepackage{makecell}

\makeatletter
\newcommand{\removelatexerror}{\let\@latex@error\@gobble}
\makeatother

\makeatletter
\newenvironment{breakablealgorithm}
  {
   \begin{center}
     \refstepcounter{algorithm}
     \hrule height.8pt depth0pt \kern2pt
     \renewcommand{\caption}[2][\relax]{
       {\raggedright\textbf{\ALG@name~\thealgorithm} ##2\par}%
       \ifx\relax##1\relax 
         \addcontentsline{loa}{algorithm}{\protect\numberline{\thealgorithm}##2}%
       \else 
         \addcontentsline{loa}{algorithm}{\protect\numberline{\thealgorithm}##1}%
       \fi
       \kern2pt\hrule\kern2pt
     }
  }{
     \kern2pt\hrule\relax
   \end{center}
  }
\makeatother

\newtheorem{myPropos}{\textbf{Proposition}}
\graphicspath{{Figures/}}
\DeclareGraphicsExtensions{.png, .eps}

\begin{document}
	\title{xURLLC-Aware Service Provisioning in Vehicular Networks: A Semantic Communication Perspective}
	\author{Le Xia,~\IEEEmembership{Graduate Student~Member,~IEEE},
				 Yao Sun,~\IEEEmembership{Senior~Member,~IEEE},
				 Dusit Niyato,~\IEEEmembership{Fellow,~IEEE},\\
				 Daquan Feng,~\IEEEmembership{Member,~IEEE},
				 Lei Feng,~\IEEEmembership{Member,~IEEE},
				 and Muhammad Ali Imran,~\IEEEmembership{Fellow,~IEEE}
	\thanks{
	
	Preliminary results of this work have been presented in part at the IEEE International Conference on Communications (ICC), 2023~\cite{2344422}. (\textit{Corresponding author: Yao Sun}.)
	
	Le Xia, Yao Sun, and Muhammad Ali Imran are with the James Watt School of Engineering, University of Glasgow, Glasgow G12 8QQ, UK (e-mail: l.xia.2@research.gla.ac.uk; \{Yao.Sun, Muhammad.Imran\}@glasgow.ac.uk).
	
	Dusit Niyato is with the School of Computer Science and Engineering, Nanyang Technological University, Singapore 639798 (e-mail: dniyato@ntu.edu.sg).
	
	Daquan Feng is with the Guangdong Province Engineering Laboratory for Digital Creative Technology, Shenzhen University, Shenzhen 518060, China (e-mail: fdquan@szu.edu.cn).
	
	Lei Feng is with the School of Computer Science, Beijing University of Posts and Telecommunications, Beijing 100876, China (e-mail: fenglei@bupt.edu.cn).
}
}	
	\maketitle
	\begin{abstract}
	Semantic communication (SemCom), as an emerging paradigm focusing on meaning delivery, has recently been considered a promising solution for the inevitable crisis of scarce communication resources.
	This trend stimulates us to explore the potential of applying SemCom to wireless vehicular networks, which normally consume a tremendous amount of resources to meet stringent reliability and latency requirements.
	Unfortunately, the unique background knowledge matching mechanism in SemCom makes it challenging to simultaneously realize efficient service provisioning for multiple users in vehicle-to-vehicle networks.
	To this end, this paper identifies and jointly addresses two fundamental problems of knowledge base construction (KBC) and vehicle service pairing (VSP) inherently existing in SemCom-enabled vehicular networks in alignment with the next-generation ultra-reliable and low-latency communication (xURLLC) requirements.
	Concretely, we first derive the knowledge matching based queuing latency specific for semantic data packets, and then formulate a latency-minimization problem subject to several KBC and VSP related reliability constraints.
	Afterward, a SemCom-empowered Service Supplying Solution (S$^{\text{4}}$) is proposed along with the theoretical analysis of its optimality guarantee and computational complexity.
	Numerical results demonstrate the superiority of S$^{\text{4}}$ in terms of average queuing latency, semantic data packet throughput, user knowledge matching degree and knowledge preference satisfaction compared with two benchmarks.
	\end{abstract}
	

	\IEEEpeerreviewmaketitle
	
	\section{Introduction}
	\IEEEPARstart{U}{biquitous} intelligence is expected to emerge in next-generation vehicular networks to accommodate diverse smart on-board applications and large-capacity vehicle-to-everything (V2X) services (e.g., multimodal artificial intelligence-generated content offered by ChatGPT or Dall-E), which poses tremendous demands on high data rates along with stringent requirements for reliability and latency~\cite{wang2018networking,she2021tutorial}.
	Correspondingly, the scarcity of available communication resources, such as bandwidth and energy, is envisioned to exacerbate to unprecedented levels and to be the most challenging problem in the near future, especially considering traditional communications-based massive vehicular networks.
	Fortunately, \textit{semantic communication} (SemCom) beyond the conventional Shannon paradigm has recently been recognized as a promising remedy for communication resource saving and transmission reliability promotion~\cite{weaver1953recent,bao2011towards,xie2021deep,strinati20216g,shi2021semantic,9797984,luo2022semantic,xia2023wiservr}, which, therefore, inspires us to investigate the potential of exploiting SemCom to perform efficient service provisioning in vehicular networks.
	
	Different from traditional communication schemes of guaranteeing the precise reception of transmitted bits~\cite{gyawali2020challenges}, the accurate delivery of semantics implied in desired messages becomes the cornerstone concept of SemCom~\cite{weaver1953recent}.
	Taking a single SemCom-enabled vehicle-to-vehicle (V2V) link as an example, a sender vehicle first leverages background knowledge relevant to source messages to filter out irrelevant content and extract core semantic features that only require fewer bits for transmission, the process of which is called semantic encoding~\cite{bao2011towards}.
	Once the receiver vehicle has the same knowledge as the sender, its local semantic interpreters are capable of accurately restoring the original meanings from the received bits, even with intolerable bit errors during data dissemination~\cite{strinati20216g}.
	This process is called semantic decoding~\cite{bao2011towards}.
	As a matter of fact, the rationale behind the growing prosperity of SemCom is due to the increasingly powerful representation and generalization abilities of semantic coding models. Particularly, powered by state-of-the-art deep learning (DL) algorithms-based semantic models, SemCom has been well explored to provision multimodal services, including text~\cite{xie2021deep}, image~\cite{shi2021semantic}, video~\cite{xia2023wiservr}, and so on.
	Consequently, SemCom is believed to significantly alleviate the resource scarcity problem, while ensuring sufficient transmission efficiency along with ultra-low semantic errors in V2V networks.
	
	Apart from many superiorities, it is worth pointing out that equivalent background knowledge should be of paramount importance to eliminate semantic ambiguity, which has led to a key concept of~\textit{knowledge base} (KB) in the realm of SemCom~\cite{strinati20216g,shi2021semantic,9797984,luo2022semantic,xia2023wiservr}.
	Specifically, a single KB is deemed a small information entity that contains background knowledge corresponding to only one particular application domain (e.g., music or sports)~\cite{strinati20216g}.\footnote{The structure of a KB can roughly cover multiple computational ontologies, facts, rules and constraints associated to a specific domain~\cite{chein2008graph}. In recent DL-driven semantic coding models, the KB is also treated as a training database serving a certain class of learning tasks~\cite{strinati20216g,luo2022semantic}. Relevant research is beyond the scope of this paper and will not be discussed in depth.}
	Since different KBs are associated with different background knowledge, holding some common KBs becomes the necessary condition to perform SemCom between two vehicles in accordance with the knowledge equivalence principle.
	Moreover, through employing differing KBs, semantic information related to different application domains can be accurately exchanged among vehicles, thereby efficiently achieving service provisioning in a way of SemCom.
	
	By reviewing the recent research advancement of SemCom, there have been some preliminary publications addressing several challenges of semantics-aware wireless networks.
	Xie~\textit{et al.}~\cite{xie2021deep} devised a Transformer-based end-to-end SemCom system to realize link-level text transmission, and then improved this system to be lightweight in~\cite{xie2020lite}.
	Proceeding as in~\cite{xie2021deep}, Yan~\textit{et al.}~\cite{yan2022resource} exploited the semantic spectral efficiency optimization-based channel assignment.
	In parallel, Xia~\textit{et al.}~\cite{9797984} developed the bit-to-message transformation and a new metric called system throughput in message for the first time to optimize resource management in SemCom-enabled cellular networks.
	Nevertheless, to the best of our knowledge, none of the existing work has ever explored the potential of applying SemCom to V2V networks in alignment with stringent latency and reliability requirements, which should be rather challenging as explained below.
	
	In full view of the novel paradigm of~\textit{SemCom-enabled vehicular networks} (SCVNs), the task lies in seeking the optimal solution to efficiently provide all participating vehicle users (VUEs) with diverse SemCom-empowered services.
	However, it is noticed that enabling the next-generation ultra-reliable and low-latency communication (xURLLC) remains indispensable, especially when pursuing adequate semantic fidelity for large-scale V2V communications.
	Uniquely, the reliability requirement originates from the aforementioned strict knowledge equivalence condition, while the latency requirement is related to varying processing efficiencies of semantic interpretation models.
	In summary, we are encountering three fundamental networking challenges in SCVN.
	\begin{itemize}
		\item \emph{Challenge 1: How to measure performance in terms of reliability and latency when introducing SemCom into vehicular networks?}
		Notice that data packets transmitted in traditional V2V communications generally consider only one type of queuing process, in which different packets have the same distributions of arrival and interpretation~\cite{guo2019resource}. However, semantic data packets in SCVN related to various SemCom-empowered services may result in different queuing and processing delays due to the different semantic interpretation models equipped on VUEs.
		Hence, it is not trivial to accurately measure and assume prior information about the latency performance in SCVN.
		Besides, given the core mechanism of semantic delivery, it should be more reasonable to characterize the reliability performance from the knowledge equivalence perspective between any two associated VUEs for V2V communications.
		All of the above constitutes the first and the main challenge.
		\item \emph{Challenge 2: How to construct appropriate KBs at each VUE for better SemCom-empowered service provisioning?}
		Considering varying practical KB sizes, personal preferences on different SemCom services, and the limited vehicular storage capacities, there is a pressing need to devise an optimal~\textit{knowledge base construction} (KBC) policy that is not only proactive but also collaborative for all VUEs to construct their respective appropriate KBs for better service provisioning.
		Notably, this challenge is inevitable in xURLLC-aware SCVN, since the remote KB access approach (i.e., the approach that each VUE remotely accesses its required KBs via RSUs, Cloud servers or core networks) can incur intolerable communication overhead and transmission latency.
		Therefore, the local and distributed KBC approach should be more applicable for each VUE to well perform SemCom.
		\item \emph{Challenge 3: How to select the best vehicle node for each VUE from multiple candidate neighbors to optimize service provisioning related overall network performance?}
		As mentioned earlier, selecting vehicle pairs for realizing service provisioning is very much distinct in SCVN due to the knowledge equivalence condition.
		Combined with different KBs constructed at numerous VUEs and unstable wireless link quality, it can be challenging to well solve the service provisioning-driven vehicle pairing problem, namely~\textit{vehicle service pairing} (VSP).
	\end{itemize}
	
	In line with the above, it is particularly worthwhile to note that challenges 2 and 3 are closely coupled, which makes it indispensable to jointly seek the optimal KBC and VSP policy for all VUEs to meet the xURLLC requirements.
	Moreover, efficient SemCom-empowered service provisioning is excepted after addressing the three challenges to yield a bunch of benefits in SCVN, such as improving V2V information interaction efficiency, reducing data traffic congestion, and ensuring high-quality vehicular services.
	
	In this paper, we propose a novel SemCom-empowered Service Supplying Solution (S$^{\text{4}}$) in SCVN with the awareness of meeting the xURLLC requirements.
	Both theoretical analysis and numerical results demonstrate the performance superiority of S$^{\text{4}}$ in terms of average queuing latency, semantic data packet throughput, user knowledge matching degree, and user knowledge preference satisfaction compared with two different benchmarks.
    In a nutshell, our main contributions are summarized as follows:
    \begin{itemize}
		\item
		We identify two fundamental yet unique problems KBC and VSP in SCVN by fully incorporating SemCom-related characteristics with vehicular network scenarios.
		In particular, individual VUE preference for different KBs is considered in the KBC, while the VSP of two adjacent VUEs takes into account the strict matching requirement between their respective constructed KBs.
		\item
		We theoretically derive the KB matching based queuing latency for a VUE pair in SCVN.
		Then, through carefully analyzing the unique queuing features of received semantic data packets, a joint latency-minimization problem is mathematically formulated subject to several KBC and VSP-related reliability constraints and other practical system limitations.
		\item
		We develop an efficient solution named S$^{\text{4}}$ to tackle the above optimization problem, and its optimality is theoretically proved by two propositions.
		Specifically, a primal-dual problem transformation method is first exploited in S$^{\text{4}}$ to obtain the corresponding Lagrange dual problem, followed by a two-stage method dedicated to solving multiple subproblems with a low computational complexity.
		Given the dual variables in each iteration, the first stage is to obtain the optimal KBC sub-policy for each potential VUE pair, whereby the second stage is able to finalize the optimal solutions of KBC and VSP for all VUEs in SCVN.
	\end{itemize}
    
    The remainder of this paper is organized as follows.
    Section II first introduces the system model of SCVN.
    Then a joint service provisioning problem in SCVN is identified and formulated in Section III.
    In Section IV, we illustrate the proposed solution S$^{\text{4}}$.
    Numerical results are presented and discussed in Section V, followed by conclusions in Section VI.
    
	\section{System Model}
	In this section, the considered SCVN scenario is first elaborated along with the knowledge storage model and vehicle pairing model.
	Then, the knowledge matching based queuing latency for semantic packets is derived.
		
	\subsection{SCVN Scenario}
	\begin{figure}[t]
		\centering
		\includegraphics[width=0.48\textwidth]{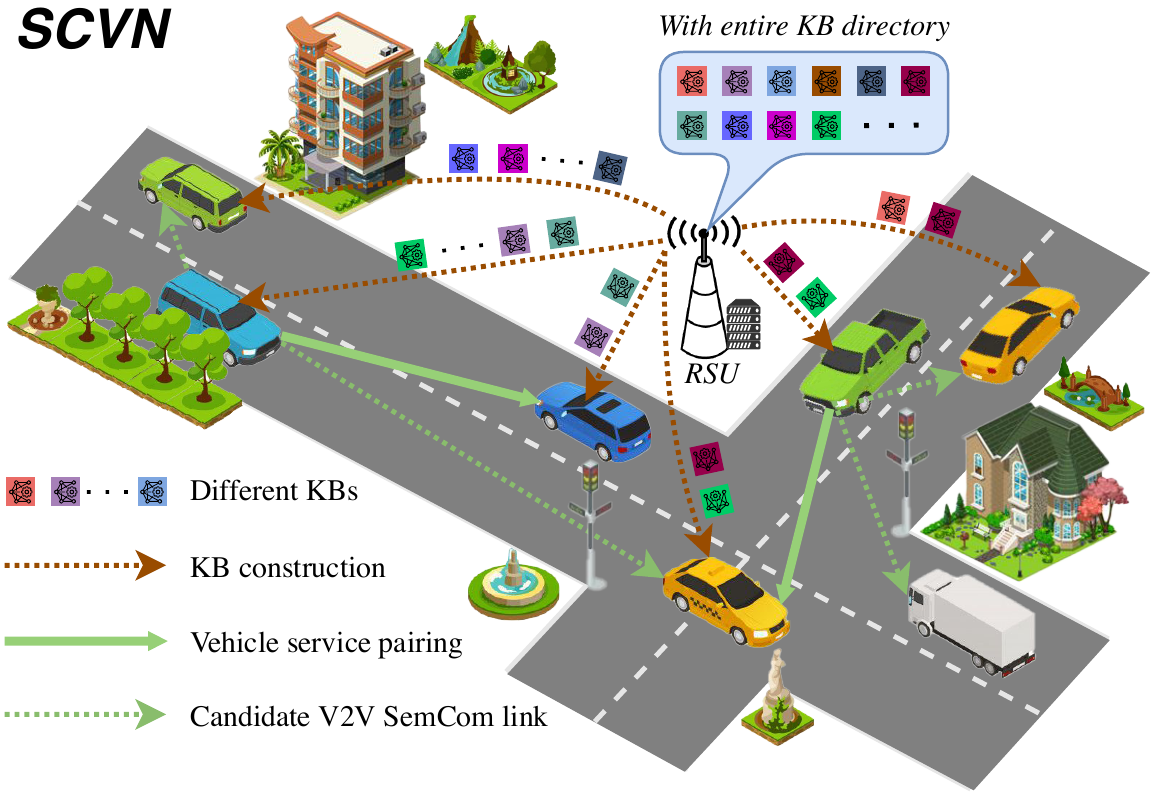} 
		\caption{The SCVN scenario with KBC and VSP.}
		\label{SCVN}
    \end{figure}
	Consider an SCVN scenario as shown in Fig.~\ref{SCVN}, the total of $V$ VUEs are distributed within the coverage of a single roadside unit (RSU), and each VUE $i \in \mathcal{V}=\left\{1, 2,\ldots,V\right\}$ is capable of providing SemCom-empowered services to others.
	For the wireless propagation model, let $\gamma_{i,j}$ denote the signal-to-interference-plus-noise ratio (SINR) experienced by the V2V link between VUE $i$ and VUE $j$ ($j \neq i$).
	Essentially, we allow VUE $i$ to communicate with VUE $j$ if their SINR value $\gamma_{i,j}$ is above a prescribed threshold $\gamma_{0}$.
	In this manner, the set of communication neighbors of VUE $i$ is defined as $\mathcal{V}_{i}=\left\{j|j \in \mathcal{V},j\neq i, \gamma_{i,j} \geqslant \gamma_{0}\right\}$, $\forall i \in \mathcal{V}$.
	Moreover, it is known that the RSU has powerful communication, computing, and storage capabilities and can provide stable communication coverage~\cite{wang2018networking}.
	Hence, in this work, let the RSU act as a semantic service controller in the SCVN to efficiently schedule and coordinate the whole SemCom-empowered service provisioning process based on the request and state information received from all participating VUEs within its coverage.
	
	\subsection{Vehicular Knowledge Storage Model}
	Due to the unique mechanism of semantic interpretation, the acquisition of necessary background knowledge is inevitable for all SemCom-enabled transceivers.
	In this work, assuming that all VUEs are able to proactively download and construct their respective required KBs from the RSU, where each VUE $i$ has a finite capacity $C_{i}$ for its local KB storage.
	Meanwhile, suppose that there is a KB library $\mathcal{K}$ with a total of $N$ differing KBs in the considered SCVN, and each requires a unique storage size $s_{n}$, $n \in \mathcal{K}=\left\{1, 2,\ldots, N\right\}$.
	Furthermore, we define a binary KBC indicator as
	\begin{equation}
		\label{alpha}
			\alpha_{i}^{n}=\left\{\begin{aligned}
			1,\quad &  \text{if KB $n$ is constructed at VUE $i$;}\\
			0,\quad &  \text{otherwise.}
		\end{aligned}
		\right.
	\end{equation}
	It is worth mentioning that the same KB cannot be constructed repeatedly at one VUE for reducing redundancy and for promoting the storage efficiency.
	
	Besides, it is noticed that different VUEs may have different preferences for these KBs corresponding to their required services, thus resulting in the diversity of KB popularity.
	Naturally, the more popular the KBs, the higher the KBC probabilities.
	Therefore, without loss of generality, we assume that the KB popularity at each VUE follows Zipf distribution~\cite{piantadosi2014zipf}.\footnote{Other known probability distributions can also be adopted without changing the remaining modeling and solution.}
	Hence, the probability of VUE $i$ requesting its desired KB $n$-based services (generating the corresponding semantic data packets) is $p_{i}^{n}=\left(r_{i}^{n}\right)^{-\xi_{i}}/\sum_{e \in \mathcal{K}}e^{-\xi_{i}}, \forall (i,n) \in \mathcal{V}\times \mathcal{K}$, where $\xi_{i}$ ($\xi_{i}\geqslant 0$) is the skewness of the Zipf distribution, and $r_{i}^{n}$ is the popularity rank of KB $n$ at VUE $i$.\footnote{The KB popularity rank of each VUE can be estimated based on its historical messaging records, which will not be discussed in this paper.}
	Based on $p_{i}^{n}$, we specially develop a KBC-related metric $\eta_{i}$, namely~\textit{knowledge preference satisfaction}, to measure the satisfaction degree of VUE $i$ constructing its interested KBs as
	\begin{equation}
		\eta_{i}=\sum_{n \in \mathcal{K}}\alpha_{i}^{n}p_{i}^{n}.
	\end{equation}
	It is further required that $\eta_{i} \geqslant \eta_{0}$, where $\eta_{0}$ is the unified minimum threshold that needs to be achieved at each VUE.
	
	\subsection{Vehicle Pairing Model for SemCom}	
	Apart from equipping with suitable KBs, it is also essential for each VUE to select an appropriate VUE from its neighbors for SemCom-empowered service provisioning.
	It is worthwhile to re-emphasize that the necessary condition for performing SemCom is that the two VUEs (transmitter and receiver) hold common KBs.
	Moreover, the single association is required for all VUEs in the SCVN for practical purposes, i.e., each VUE can be paired with only one (another) VUE at a time.
	
	Let $\beta_{i\looparrowright j}$ denote the binary VSP indicator for a VUE $i$-VUE $j$ pair (suppose that VUE $i$ is the sender and VUE $j$ is the receiver), where
	\begin{equation}
		\label{alpha}
			\beta_{i\looparrowright j}=\left\{\begin{aligned}
			1,\quad &  \text{if VUE $i$ is associated with VUE $j$;}\\
			0,\quad &  \text{otherwise.}
		\end{aligned}
		\right.
	\end{equation}
	Note that the presented communication performance (such as latency, reliability, and throughput) should be different when swapping the roles of sender and receiver in the same VUE pair, since the KBs utilized for SemCom are determined by the sender's preference.
	For this reason, we use the notation $\looparrowright$ here as an auxiliary illustration to specify the roles of the sender and receiver in each VUE pair.
	
	\subsection{Knowledge Matching Based Queuing Model}
	\begin{figure}[t]
		\centering
		\includegraphics[width=0.48\textwidth]{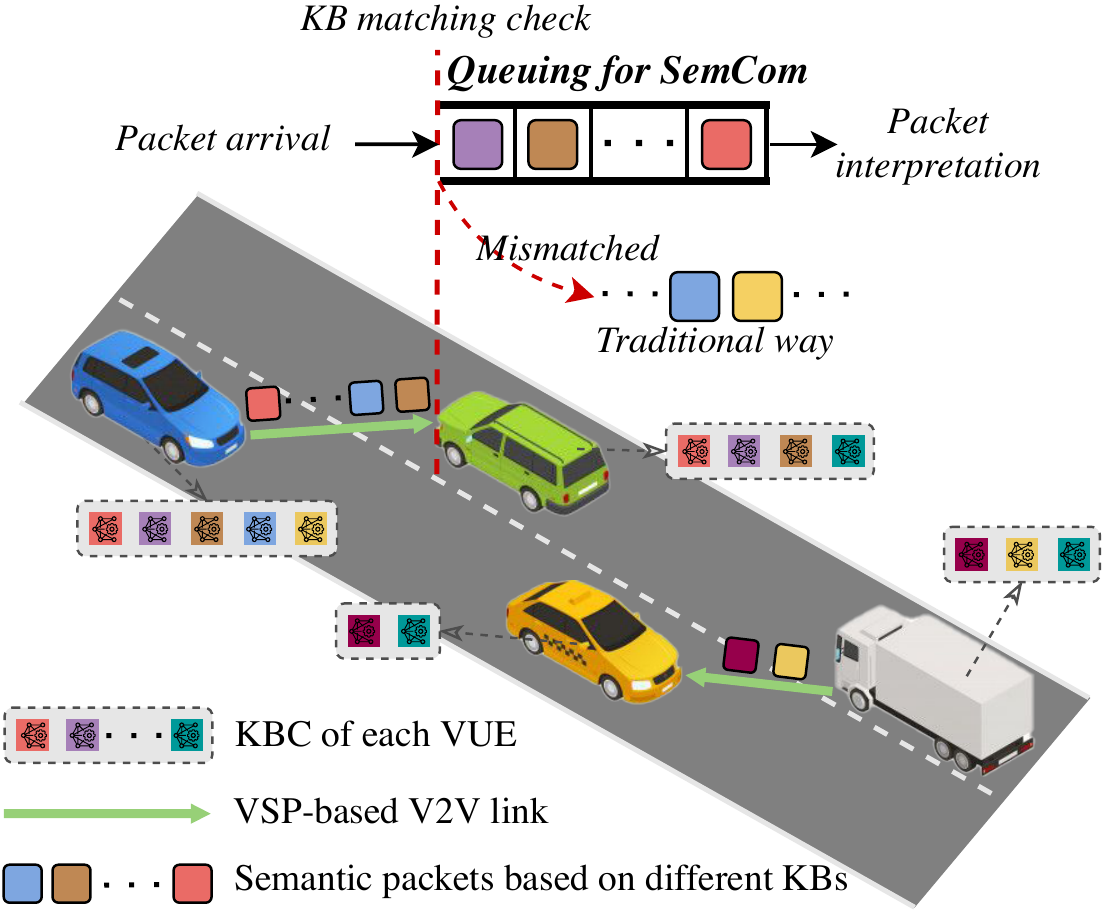} 
		\caption{The knowledge matching based queuing model for semantic data packets transmitted between VUEs in the SCVN.}
		\label{Queuing}
    \end{figure}
	As depicted in Fig.~\ref{Queuing}, the knowledge matching based semantic packet queuing delay is employed as the latency metric of SemCom, to characterize the average sojourn time of semantic data packets in the receiver VUE's queue buffer (following the first-come first-serve rule).
	For better illustration, three major differences between SemCom-based and traditional communication-based queuing models are listed below:
	1) Each semantic data packet is associated with a specific service type, i.e., a certain KB;
	2) Semantic data packets generated based on different KBs can co-exist in the queue, and have independent average arrival rate and interpretation time;
	3) Not all semantic data packets arriving at the receiver VUE are always allowed to enter its queue, as some of them may mismatch the KBs currently held, rendering these packets uninterpretable.
	To avoid pointless queuing, these mismatched packets may have to choose traditional communication channels for information transfer, and will not be counted in the arrival process of the queue.
	
	To preserve generality, we first suppose a Poisson data arrival process with average rate $\lambda_{i}^{n}=\lambda_{i}p_{i}^{n}$ for a sender VUE $i$ to account for its local semantic packet generation based on KB $n$, where $\lambda_{i}$ is the total arrival rate of all semantic packets at VUE $i$.
	In line with this, we can obtain the overall arrival rate of semantic packets from sender VUE $i$ to receiver VUE $j$ as $\sum_{n \in \mathcal{K}}\alpha_{i}^{n}\lambda_{i}^{n}$, and the effective arrival rate of semantic packets (i.e., these KB-matched semantic packets) in the queue is given as $\sum_{n \in \mathcal{K}}\alpha_{i}^{n}\alpha_{j}^{n}\lambda_{i}^{n}$, thereby the arrival rate of mismatched semantic packets should be the value of the former minus the latter, that is, $\sum_{n \in \mathcal{K}}\alpha_{i}^{n}\left(1-\alpha_{j}^{n}\right)\lambda_{i}^{n}$.
	Herein, denoting the ratio of the arrival rate of mismatched semantic packets to the arrival rate of all received semantic packets at VUE $i$-VUE $j$ pair as $\theta_{i\looparrowright j}$, namely~\textit{knowledge mismatch degree}, which is explicitly calculated by
	\begin{equation}
		\label{kmd}
		\begin{aligned}
			\theta_{i\looparrowright j}=\frac{\sum_{n \in \mathcal{K}}\alpha_{i}^{n}\left(1-\alpha_{j}^{n}\right)\lambda_{i}^{n}}{\sum_{n \in \mathcal{K}}\alpha_{i}^{n}\lambda_{i}^{n}}.
		\end{aligned}
	\end{equation}
		
	In parallel, let a random variable $I_{j}^{n}$ denote the Markovian interpretation time~\cite{lavee2009understanding} required by KB $n$-based packets at VUE $j$ with mean $1/\mu_{j}^{n}$, which is determined by the computing capability of the vehicle and the type of the desired KB.
	However, since multiple packets based on different KBs are allowed to queue at the same time, it is seen that the interpretation time distribution for a receiver VUE should be treated as a general distribution~\cite{ross2014introduction}.
	If further taking into account the KB popularity, we can calculate the ratio of the amount of KB $n$-based packets to the total packets in the VUE $i$-VUE $j$ pair's queue by $\epsilon_{i\looparrowright j}^{n}=p_{i}^{n}/\sum_{f \in \mathcal{K}}\alpha_{i}^{f}\alpha_{j}^{f}p_{i}^{f}$.
	With the independence among packets based on different KBs, the interpretation time required by each packet in the queue is now expressed as $W_{i\looparrowright j}=\sum_{n \in \mathcal{K}}\alpha_{i}^{n}\alpha_{j}^{n}\epsilon_{i\looparrowright j}^{n}I_{j}^{n}$.
	
	Since the Markovian arrival process leads to the correlated packet arrivals while the service pattern of packets obeys a general distribution, the queue of each VUE pair in the SCVN can be modeled as an M/G/1 system, which has been widely used to model data traffic in wireless networks.
	According to the~\textit{Pollaczek-Khintchine formula}~\cite{pollaczek1930aufgabe}, the average queuing latency for the VUE $i$-VUE $j$ pair, denoted as $\delta_{i\looparrowright j}$, is determined as follows\footnote{In order to guarantee the steady-state of the queuing system, a condition of $\lambda_{i\looparrowright j}^{\mathit{eff}}\mathds{E}\left[W_{i\looparrowright j}\right]<1$ must be satisfied before proceeding~\cite{ross2014introduction}. In this work, we assume that the packet interpretation rate is larger than the packet arrival rate to make the queuing latency finite and thus solvable.}
	\begin{equation}
		\delta_{i\looparrowright j}=\frac{\lambda_{i\looparrowright j}^{\mathit{eff}}\cdot \left(\mathds{E}^{2}\left[W_{i\looparrowright j}\right]+\mathit{Var}\left(W_{i\looparrowright j}\right)\right)}{2\left(1-\lambda_{i\looparrowright j}^{\mathit{eff}}\cdot \mathds{E}\left[W_{i\looparrowright j}\right]\right)}.\label{PK}
	\end{equation}
	On this basis, again leveraging the independence of $I_{j}^{n}$ over $n$, we can then obtain the expectation of the interpretation time for all semantic data packets in the queue by
	\begin{equation}
		\begin{aligned}
		\mathds{E}\left[W_{i\looparrowright j}\right]&=\sum_{n \in \mathcal{K}}\alpha_{i}^{n}\alpha_{j}^{n}\epsilon_{i\looparrowright j}^{n}\mathds{E}\left[I_{j}^{n}\right]=\sum_{n \in \mathcal{K}}\frac{\alpha_{i}^{n}\alpha_{j}^{n}\epsilon_{i\looparrowright j}^{n}}{\mu_{j}^{n}},\label{PK1}
		\end{aligned}
	\end{equation}
	and the variance of $W_{i\looparrowright j}$ is given by
	\begin{equation}
		\begin{aligned}
		\mathit{Var}\left(W_{i\looparrowright j}\right)&=\sum_{n \in \mathcal{K}}\alpha_{i}^{n}\alpha_{j}^{n}\left(\epsilon_{i\looparrowright j}^{n}\right)^{2}\mathit{Var}\left[I_{j}^{n}\right]\\
		&=\sum_{n \in \mathcal{K}}\alpha_{i}^{n}\alpha_{j}^{n}\left(\frac{\epsilon_{i\looparrowright j}^{n}}{\mu_{j}^{n}}\right)^{2}.\label{PK2}
		\end{aligned}
	\end{equation}
	Based on~(\ref{PK1}) and (\ref{PK2}), $\delta_{i\looparrowright j}$ in~(\ref{PK}) can be rewritten in~(\ref{Qdelay}), as shown at the bottom of this page.
	\begin{figure*}[hb]
		\centering
		\hrulefill
		\begin{equation}
			\delta_{i\looparrowright j}=\frac{\left[\left(\sum_{n \in \mathcal{K}}\alpha_{i}^{n}\alpha_{j}^{n}\frac{p_{i}^{n}/\mu_{j}^{n}}{\sum_{f \in \mathcal{K}}\alpha_{i}^{f}\alpha_{j}^{f}p_{i}^{f}}\right)^{2}+\sum_{n \in \mathcal{K}}\alpha_{i}^{n}\alpha_{j}^{n}\left(\frac{p_{i}^{n}/\mu_{j}^{n}}{\sum_{f \in \mathcal{K}}\alpha_{i}^{f}\alpha_{j}^{f}p_{i}^{f}}\right)^{2}\right]\cdot \left(\sum_{n \in \mathcal{K}}\alpha_{i}^{n}\alpha_{j}^{n}\lambda_{i}^{n}\right)}{2\left[1-\left(\sum_{n \in \mathcal{K}}\alpha_{i}^{n}\alpha_{j}^{n}\lambda_{i}^{n}\right)\cdot \left(\sum_{n \in \mathcal{K}}\alpha_{i}^{n}\alpha_{j}^{n}\frac{p_{i}^{n}/\mu_{j}^{n}}{\sum_{f \in \mathcal{K}}\alpha_{i}^{f}\alpha_{j}^{f}p_{i}^{f}}\right)\right]}.\label{Qdelay}
		\end{equation}
	\end{figure*}
	
	\section{Problem Formulation}
	In line with the xURLLC requirements, it is of paramount importance to achieve the optimality of the overall queuing delay in the SCVN, while being subject to several SemCom-relevant reliability requirements as well as practical system constraints.
	To that end, we identify and formulate a latency-minimization problem in a joint optimization manner of the KBC indicator $\alpha_{i}^{n}$ and the VSP indicator $\beta_{i\looparrowright j}$.
	For ease of illustration, we define a matrix $\bm{\alpha}=\left\{\alpha_{i}^{n}|i \in \mathcal{V},n \in \mathcal{K}\right\}$ and a matrix $\bm{\beta}=\left\{\beta_{i\looparrowright j}|i \in \mathcal{V},j \in \mathcal{V}_{i}\right\}$ consisting of all binary variables of KBC and VSP, respectively.
	On the basis of these, our joint optimization problem is formulated as follows:
	\setcounter{equation}{8}
    \begin{align}
	\mathbf{P0}:\ \min_{\bm{\alpha},\bm{\beta}} \quad & \sum_{i \in \mathcal{V}}\sum_{j \in \mathcal{V}_{i}}\beta_{i\looparrowright j}\delta_{i\looparrowright j} ~\label{P0}\\
	{\rm s.t.} \quad & \sum_{n \in \mathcal{K}} \alpha_{i}^{n}\cdot s_{n}\leqslant C_{i},\  \forall i\in \mathcal{V},\tag{\ref{P0}a}\\
	& \eta_{i}\geqslant \eta_{0},\ \forall i \in \mathcal{V},\tag{\ref{P0}b}\\
	&\sum_{j \in \mathcal{V}_{i}}\beta_{i\looparrowright j}= 1,\ \forall i \in \mathcal{V},\tag{\ref{P0}c}\\
	& \beta_{i\looparrowright j}=\beta_{j\looparrowright i},\ \forall \left( i,j\right) \in \mathcal{V}\times \mathcal{V}_{i},\tag{\ref{P0}d}\\
	& \sum_{j \in \mathcal{V}_{i}}\beta_{i\looparrowright j}\theta_{i\looparrowright j}\leqslant\theta_{0},\ \forall i \in \mathcal{V},\tag{\ref{P0}e}\\
	& \alpha_{i}^{n}\in \left\{ 0,1\right\},\ \forall \left( i,n\right) \in \mathcal{V}\times \mathcal{K},\tag{\ref{P0}f}\\
	& \beta_{i\looparrowright j}\in \left\{ 0,1\right\},\ \forall \left( i,j\right) \in \mathcal{V}\times \mathcal{V}_{i}.\tag{\ref{P0}g}
	\end{align}
	Constraint (\ref{P0}a) ensures that the total size of KBs constructed at each vehicle cannot exceed its maximum storage capacity, while constraint (\ref{P0}b) corresponds to the aforementioned knowledge preference satisfaction requirement for each VUE.
	Constraints (\ref{P0}c) and (\ref{P0}d) mathematically model the single-association requirement of VUEs.
	Constraint (\ref{P0}e) represents that the knowledge mismatch degree of each VUE pair should not be over the threshold $\theta_{0}$, which guarantees sufficiently high reliability of semantic information delivery.
    Constraints (\ref{P0}f) and (\ref{P0}g) characterize the binary properties of $\bm{\alpha}$ and $\bm{\beta}$, respectively.
    
    Carefully examining $\mathbf{P0}$, it is seen that addressing this problem is rather challenging due to several inevitable mathematical obstacles.
    First of all, $\mathbf{P0}$ is an NP-hard optimization problem as demonstrated below.
    Consider a special case of $\mathbf{P0}$ where all $\bm{\beta}$-related constraints are satisfied.
    In this case, constraints (\ref{P0}c)-(\ref{P0}e) and (\ref{P0}g) can all be removed, and the primal problem degenerates into a classical $0$-$1$ multi-knapsack problem that is known to be NP-hard~\cite{kellerer2004multidimensional}.
    Hence, $\mathbf{P0}$ is also NP-hard.
    Another nontrivial point originates from the complicated objective function, which prevents us from using the conventional two-step solution (i.e., relaxation and recovery) to approach optimality.
    In more detail, the problem after relaxing $\bm{\alpha}$ and $\bm{\beta}$ should still be a nonconvex optimization problem owing to the non-convexity preserved in the objective function (\ref{P0}) and constraint (\ref{P0}e).
  	Therefore, a severe performance penalty will be incurred from the procedure of integer recovery due to the huge performance compromise on solving the nonconvex problem for relaxed variables~\cite{tuy1998convex,papadimitriou1998combinatorial,burer2012non}.
  	In view of the above mathematical challenges, we propose an efficient solution S$^{\text{4}}$ in the subsequent section to solve $\mathbf{P0}$ and obtain the joint optimal KBC and VSP solution.
  	
  	\section{Proposed SemCom-Empowered Service Supplying Solution (S$^{\text{4}}$)}
\begin{figure}[h]
		\centering
		\includegraphics[width=0.48\textwidth]{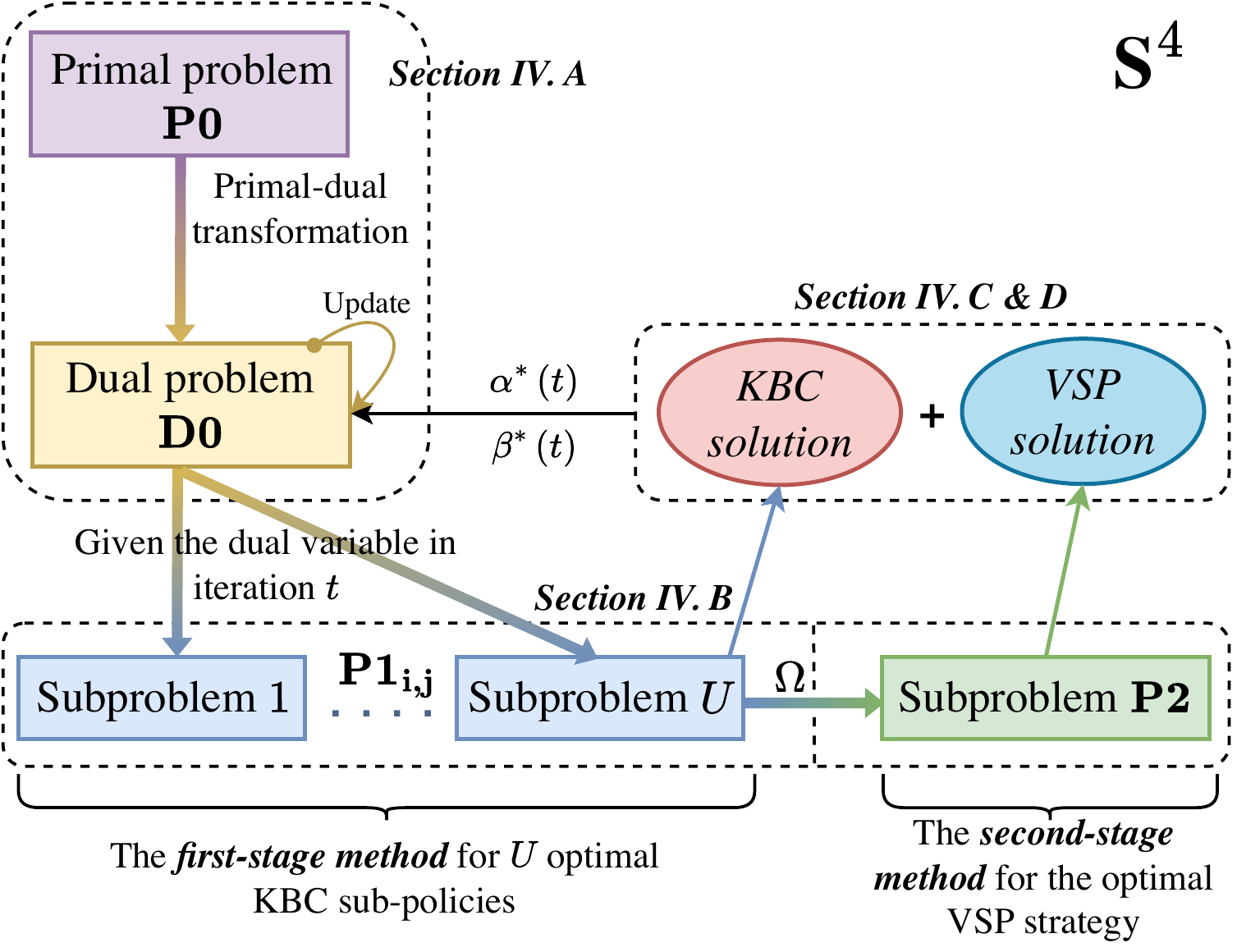} 
		\caption{Illustration of the proposed solution S$^{\text{4}}$.}
		\label{Solutionfig}
    \end{figure}

  	In this section, we illustrate how to design our proposed solution S$^{\text{4}}$ to cope with the SemCom-empowered service provisioning problem $\mathbf{P0}$ in vehicular networks.
 	As depicted in Fig.~\ref{Solutionfig}, a Lagrange dual method is first leveraged to eliminate the cross-term constraints in $\mathbf{P0}$ with a corresponding dual optimization problem transformed (referring to $\mathbf{D0}$ in Section IV.A).
 	Then given the dual variable in each iteration, we dedicatedly develop a two-stage method to determine $\bm{\alpha}$ and $\bm{\beta}$ for the dual problem, where the optimality will be theoretically proved in Section IV.B.
 	Specifically, in the first stage, we subtly construct $U$ ($U=\left(\sum_{i \in \mathcal{V}}\left|\mathcal{V}_{i}\right|\right)/2$) subproblems (referring to $\mathbf{P1}_{i,j},\forall \left( i,j\right) \in \mathcal{V}\times \mathcal{V}_{i}, j>i$), each of which aims to independently seek the optimal KBC sub-policy (with respect to only $\alpha_{i}^{n}$ and $\alpha_{j}^{n}$) for each individual VUE pair (as detailed in Section IV.C).
 	After solving all the $U$ subproblems, the optimal coefficient matrix $\bm{\Omega}$ is obtained for all potential VUE pairs in the SCVN, by which we further construct a new subproblem (referring to $\mathbf{P2}$) in the second stage to find the optimal VSP strategy for $\bm{\beta}$ (as detailed in Section IV.D).
 	In the end, we present the workflow of S$^{\text{4}}$ along with its complexity analysis in Section IV.E.
  	
  	\subsection{Primal-Dual Problem Transformation}
    We first incorporate constraint (\ref{P0}e) into the objective function (\ref{P0}) by associating a Lagrange multiplier $\bm{\tau}=\{\tau_{i}|i \in \mathcal{V}\}$.
    That way, the associated Lagrange function is obtained by
    \begin{equation}
		\begin{aligned}
			&L\left(\bm{\alpha},\bm{\beta},\bm{\tau}\right)\\
			&\ \quad = \sum_{i \in \mathcal{V}}\sum_{j \in \mathcal{V}_{i}}\beta_{i\looparrowright j}\delta_{i\looparrowright j}+\sum_{i \in \mathcal{V}}\tau_{i}\left(\sum_{j \in \mathcal{V}_{i}}\beta_{i\looparrowright j}\theta_{i\looparrowright j}-\theta_{0}\right)\\
			&\ \quad = \sum_{i \in \mathcal{V}}\sum_{j \in \mathcal{V}_{i}}\beta_{i\looparrowright j}\left(\delta_{i\looparrowright j}+\tau_{i}\theta_{i\looparrowright j}\right)-\theta_{0}\sum_{i \in \mathcal{V}}\tau_{i}\\
			&\ \quad \triangleq \widetilde{L}_{\bm{\tau}}(\bm{\alpha},\bm{\beta})-\theta_{0}\sum_{i \in \mathcal{V}}\tau_{i},\label{Lagrangian}
		\end{aligned}
	\end{equation}
    where $\widetilde{L}_{\bm{\tau}}(\bm{\alpha},\bm{\beta})$ is defined for expression brevity.
    Then, the Lagrange dual problem of $\mathbf{P0}$ should be formulated as
     \begin{align}
			\mathbf{D0}:\ \max_{\bm{\tau}} \quad & D\left(\bm{\tau}\right)=g_{\bm{\alpha},\bm{\beta}}\left(\bm{\tau}\right)-\theta_{0}\sum_{i \in \mathcal{V}}\tau_{i}~\label{D}\\
			{\rm s.t.} \quad & \tau_{i}\geqslant 0,\ \forall i \in \mathcal{V},\tag{\ref{D}a}
	\end{align}
	where we have
	\begin{equation}
		\label{Dual}
			\begin{aligned}
			g_{\bm{\alpha},\bm{\beta}}\left(\bm{\tau}\right) \ &= \ \inf_{\bm{\alpha},\bm{\beta}} \ \widetilde{L}_{\bm{\tau}}(\bm{\alpha},\bm{\beta})\\
			{\rm s.t.} \ & \ \text{(\ref{P0}a)}-\text{(\ref{P0}d)}, \text{(\ref{P0}f)}, \text{(\ref{P0}g)}.
			\end{aligned}
	\end{equation}
	Notably, the optimality of the convex problem $\mathbf{D0}$ gives at least the best lower bound of $\mathbf{P0}$, even if $\mathbf{P0}$ is nonconvex, according to the duality property~\cite{boyd2004convex}.
	Hence, our focus now naturally shifts to seeking the optimal solution to $\mathbf{D0}$. 
	
	Given the initial dual variable $\bm{\tau}$, we can solve problem~(\ref{Dual}) in the first place to find the optimal solution $\bm{\alpha}$ and $\bm{\beta}$, the details of which will be presented in the next subsection.
	After that, a subgradient method is employed for updating $\bm{\tau}$ to solve $\mathbf{D0}$ in an iterative fashion, as shown in Fig.~\ref{Solutionfig}.
	Specifically, the partial derivatives with respect to $\bm{\tau}$ in the objective function $D\left(\bm{\tau}\right)$ are set as the subgradient direction in each iteration.
	Now suppose that in a certain iteration, say iteration $t$, each dual variable $\tau_{i}(t)$ ($i \in \mathcal{V}$) is updated as
	\begin{equation}
		\tau_{i}(t+1)=\left[\tau_{i}(t)-\nu(t)\cdot \left(\theta_{0}-\sum_{j \in \mathcal{V}_{i}}\beta_{i\looparrowright j}\left(t\right)\theta_{i\looparrowright j}\left(t\right)\right)\right]^{+}.\label{lagupdate}
	\end{equation}
	The operator $[\cdot]^{+}$ here is to output the maximum value between its argument and zero, ensuring that $\bm{\tau}$ must be non-negative as constrained in~(\ref{D}a).
	$\nu(t)$ is the stepsize in iteration $t$ and generally, the convergence of the subgradient descent method can be ensured with the proper stepsize~\cite{boyd2003subgradient}.
	
	\subsection{Two-Stage Method Based on KBC and VSP}
	As discussed before, for a given $\bm{\tau}$ in each iteration, the optimal $\bm{\alpha}$ and $\bm{\beta}$ need to be determined by solving problem~(\ref{Dual}).
    However, solving such a problem is still rather tricky due to the mathematical inseparability of $\bm{\alpha}$ and $\bm{\beta}$ in the highly complex objective function $\widetilde{L}_{\bm{\tau}}(\bm{\alpha},\bm{\beta})$.
    To this end, we propose a two-stage method to obtain the exactly optimal $\bm{\alpha}$ and $\bm{\beta}$ with a low computational complexity.
	
	In the first stage, we focus on multiple independent KB construction subproblems, each corresponding to a potential VUE pair in the SCVN.
	In particular, here the performances of the VUE $i$-VUE $j$ pair (i.e., the sender VUE $i$ and the receiver VUE $j$) and the VUE $j$-VUE $i$ pair (i.e., the sender VUE $j$ and the receiver VUE $i$) need to be considered together, and for ease of distinction, we refer to the two as a~\textit{VUE $i,j$ pair}, $\forall \left( i,j\right) \in \mathcal{V}\times \mathcal{V}_{i}, j>i$.
	In other words, for any KBC subproblem, we have $\beta_{i\looparrowright j}=\beta_{j\looparrowright i}=1$ in $\widetilde{L}_{\bm{\tau}}(\bm{\alpha},\bm{\beta})$ corresponding to a given VUE $i,j$ pair, while all other VUE pairs are not considered.
	Therefore, different KBC subproblems can be solved independently, and in this way, let
	\begin{equation}
			\omega_{i,j}=\left(\delta_{i\looparrowright j}+\tau_{i}\theta_{i\looparrowright j}\right)+\left(\delta_{j\looparrowright i}+\tau_{j}\theta_{j\looparrowright i}\right).~\label{VUEpaircost}
	\end{equation}
	Obviously, we have $\omega_{i,j}=\omega_{j,i}$, thus only one case of $j>i$ needs to be investigated for each potential VUE $i,j$ pair.

	In this context, we now construct $U=\left(\sum_{i \in \mathcal{V}}\left|\mathcal{V}_{i}\right|\right)/2$ subproblems, each of which is denoted as $\mathbf{P1}_{i,j}$ to seek the optimal KBC sub-policy only for an individual VUE $i,j$ pair.
	Herein, it is worth pointing out that the optimal KBC solution to problem~(\ref{Dual}) cannot be achieved by simply combining the obtained sub-policies of these $\mathbf{P1}_{i,j}$, but these sub-policies will be used to construct the subsequent VSP subproblem to finalize the joint optimal solution of $\bm{\alpha}$ and $\bm{\beta}$ for~(\ref{Dual}).
	Given the dual variable $\bm{\tau}$ in each iteration,\footnote{For simplicity, we omit $\bm{\tau}$ from all notations associated with $\mathbf{P1}_{i,j}$ and $\mathbf{P2}$ in this paper.} $\mathbf{P1}_{i,j}$ becomes
	\begin{align}
		\mathbf{P1}_{i,j}:\ \min_{\left\{\alpha_{i}^{n}\right\},\left\{\alpha_{j}^{n}\right\}}\quad &\omega_{i,j}~\label{P1u}\\
		{\rm s.t.} \quad \quad \ & \sum_{n \in \mathcal{K}} \alpha_{i}^{n}\cdot s_{n}\leqslant C_{i},\tag{\ref{P1u}a}\\
		&\sum_{n \in \mathcal{K}} \alpha_{j}^{n}\cdot s_{n}\leqslant C_{j},\tag{\ref{P1u}b}\\
		&\eta_{i}\geqslant\eta_{0},\ \eta_{j}\geqslant\eta_{0},\tag{\ref{P1u}c}\\
		& \alpha_{i}^{n}\in \left\{ 0,1\right\}, \alpha_{j}^{n}\in \left\{ 0,1\right\},\ \forall n \in \mathcal{K}.\tag{\ref{P1u}d}
	\end{align}
	By solving $\mathbf{P1}_{i,j}$,\footnote{The solution details of $\mathbf{P1}_{i,j}$ as well as $\mathbf{P2}$ will be introduced in the subsequent Subsection C and D, respectively.} we can obtain the optimal KBC sub-policies for VUE $i$ (denoted as $\bm{\alpha}^{*}_{i_{(j)}}$) and VUE $j$ (denoted as $\bm{\alpha}^{*}_{j_{(i)}}$),\footnote{For auxiliary illustration, we use $(\cdot)$ in the subscript to specify the VUE pair attribute (relation) for each VUE's KBC sub-policy obtained from $\mathbf{P1}_{i,j}$.} corresponding to the individual VUE $i,j$ pair.
	The following proposition explicitly shows how the sub-policy of $\mathbf{P1}_{i,j}$ correlates to the solution of problem~(\ref{Dual}).
	
	\begin{myPropos}
		Let $\bm{\alpha}^{*}=\left[\bm{\alpha}^{*}_{1}, \bm{\alpha}^{*}_{2}, \cdots,\bm{\alpha}^{*}_{V}\right]^{T}$ be the optimal KBC solution to the problem in~(\ref{Dual}) given the dual variable $\bm{\tau}$, where $\bm{\alpha}^{*}_{i}$ represents the optimal KBC policy of VUE $i$.
		Then we have $\forall i \in \mathcal{V}$, $\exists j \in \mathcal{V}_{i}$, such that $\bm{\alpha}^{*}_{i_{(j)}}=\bm{\alpha}^{*}_{i}$.
	\end{myPropos}
	\begin{IEEEproof}
			Please see Appendix A.
	\end{IEEEproof}
	
	From Proposition 1, it is observed that the optimal KBC policy of each VUE can be found by solving a certain $\mathbf{P1}_{i,j}$.
	Hence, considering the single-association requirement of V2V pairing, the optimal VSP strategy becomes the only key to finalize the optimal solution to~(\ref{Dual}).
	To achieve this, we first obtain the optimal coefficient matrix for $\bm{\beta}$ in~(\ref{Dual}) to account for all VSP possibilities.
	By calculating optimum $\omega_{i,j}$ (denoted as $\omega_{i,j}^{*},\forall \left( i,j\right) \in \mathcal{V}\times \mathcal{V}_{i}$) in $\mathbf{P1}_{i,j}$, the optimal coefficient matrix is formed as
	\begin{equation}
		\label{optimatrix}
		\bm{\Omega}=\left[
			\begin{matrix}
				+\infty &\omega_{1,2}^{*}&\omega_{1,3}^{*}&\cdots & \omega_{1,V}^{*}\\
				\omega_{2,1}^{*} &+\infty&\omega_{2,3}^{*}&\cdots & \omega_{2,V}^{*} \\
				\omega_{3,1}^{*} &\omega_{3,2}^{*}&+\infty&\cdots & \omega_{3,V}^{*} \\
				\vdots & \vdots & \vdots &\ddots & \vdots\\
				\omega_{V,1}^{*} & \omega_{V,2}^{*} & \omega_{V,3}^{*}& \cdots & +\infty
			\end{matrix}
			\right].
	\end{equation}
	$\bm{\Omega}$ is a $V\times V$ symmetric matrix where $\omega_{i,j}^{*}=\omega_{j,i}^{*}$, and all elements on its main diagonal are set to $+\infty$ to indicate the fact that a VUE cannot communicate with itself, i.e., $j\neq i$.
	Besides, note that some $\omega_{i,j}^{*}$s in $\bm{\Omega}$ also have a value $+\infty$ if VUE $j$ is not the direct neighbor of VUE $i$, i.e., $j \notin \mathcal{V}_{i}$.
	
	Next, we concentrate upon the optimal vehicle service pairing strategy by constructing a new subproblem in the second stage.
	In line with the objective $\widetilde{L}_{\bm{\tau}}(\bm{\alpha},\bm{\beta})$ and $\bm{\beta}$-related constraints in~(\ref{Dual}), the VSP subproblem is written as
	\begin{align}
		\mathbf{P2}:\ \min_{\bm{\beta}}\quad &\frac{1}{2}\sum_{i \in \mathcal{V}}\sum_{j \in \mathcal{V}_{i}}\beta_{i\looparrowright j}\omega_{i,j}^{*}~\label{P2}\\
		{\rm s.t.} \quad & \text{(\ref{P0}c)},\text{(\ref{P0}d)}, \text{(\ref{P0}g)}.\tag{\ref{P2}a}
	\end{align}
	Given any $\bm{\tau}$, the optimal VSP strategy $\bm{\beta}$ (denoted as $\bm{\beta}^{*}=\left[\beta_{1\looparrowright j^{*}_{1}},\beta_{2\looparrowright j^{*}_{2}},\cdots,\beta_{V\looparrowright j^{*}_{V}}\right]^{T}$) can be directly finalized by solving $\mathbf{P2}$, where $\beta_{i\looparrowright j^{*}_{i}}$ ($\forall i \in \mathcal{V}$) indicates that VUE $j^{*}_{i}$ is the optimal SemCom node for VUE $i$, i.e., $\beta_{i\looparrowright j^{*}_{i}}=1$.
	Afterward, we feed back the obtained $\bm{\beta}^{*}$ to $\bm{\Omega}$ to further finalize the optimal KBC policy $\bm{\alpha}^{*}$ for all VUEs.
	In the context of the solution to $\mathbf{P1}_{i,j}$, the approach to finalize $\bm{\alpha}^{*}$ can be stated more precisely as: for any $i \in \mathcal {V}$, we have $\bm{\alpha}_{i}^{*}=\bm{\alpha}^{*}_{i_{(j^{*}_{i})}}$.
	The rationale behind this is established in accordance with the following proposition.
	
	\begin{myPropos}
		Given any dual variable $\bm{\tau}$, $\left(\bm{\alpha}^{*},\bm{\beta}^{*}\right)$ is exactly the optimal solution to the problem in~(\ref{Dual}).
	\end{myPropos}
	\begin{IEEEproof}
			Please see Appendix B.
	\end{IEEEproof}
	
     From Proposition 2, it is seen that for problem~(\ref{Dual}) in each iteration, the proposed two-stage method is ensured to find the optimal solution.
     Apart from this, the optimization difficulty of each subproblem, either $\mathbf{P1}_{i,j}$ or $\mathbf{P2}$, is considered to be greatly decreased due to the reduced number of the optimization variables.
     In what follows, we will present our optimal solutions to $\mathbf{P1}_{i,j}$ and $\mathbf{P2}$, respectively.
	
	\subsection{Near-Optimal Solution to $\mathbf{P1}_{i,j}$}
	Carefully examining $\mathbf{P1}_{i,j}, \forall \left( i,j\right) \in \mathcal{V}\times \mathcal{V}_{i}, j>i$, it can be observed that $\delta_{i \looparrowright j}$ mainly makes the objective function $\omega_{i,j}$ still highly complex and nonconvex with two binary variables $\bm{\alpha}_{i}$ and $\bm{\alpha}_{j}$.
	To that end, a modified metaheuristic algorithm based on tabu search (TS) is employed here to efficiently determine a near-optimal KB construction policy for two VUEs in the VUE $i,j$ pair with the consideration of SemCom features.
	In detail, the KBC solution is illustrated as follows:
	\begin{itemize}
		\item \textit{Initial Feasible Solution Generation:}
		As the iterative search algorithm, an initial feasible solution (denoted as a $2N$-dimensional vector $\bm{\alpha}_{i,j}^{I}$) is needed as the search starting point~\cite{salhi2002defining}.
		To speed up convergence and enhance optimization performance, we heuristically adopt a KB preference and KB matching-aware approach to generate $\bm{\alpha}_{i,j}^{I}$ for a better initial solution performance.
		More concretely, first let two $N$-dimensional vectors $\bm{\alpha}_{i}^{I}$ and $\bm{\alpha}_{j}^{I}$ denote the KBC solutions of VUE $i$ and VUE $j$, respectively, with all elements being $0$ for initialization.
		Meanwhile, suppose there are two variable sets, denoted as $\hat{\mathcal{K}}$ and $\check{\mathcal{K}}$, to record KB-relevant information, where $\hat{\mathcal{K}}=\mathcal{K}$ and $\check{\mathcal{K}}=\emptyset$ are initialized.
		With these, we attempt to find a KB $n_{0}$ with the highest sum of KB preferences of the two VUEs by
		\begin{equation}
			n_{0}=\arg \max_{n\in \hat{\mathcal{K}}}\ \left(p_{i}^{n}+p_{j}^{n}\right).\label{premax}
		\end{equation}
		Then, in order to meet the knowledge mismatch requirement, the values of both $\bm{\alpha}_{i}^{I}$ and $\bm{\alpha}_{j}^{I}$ are updated as
		\begin{equation}
			\alpha_{i}^{n_{0}}=1 \quad \text{and}\quad \alpha_{j}^{n_{0}}=1.\label{premax2}
		\end{equation}
		Next, we let $\hat{\mathcal{K}}=\hat{\mathcal{K}}\backslash n_{0}$ and $\check{\mathcal{K}}=\check{\mathcal{K}}\cup\{n_{0}\}$, and then repeat the two procedures in~(\ref{premax}) and~(\ref{premax2}) until both VUEs satisfy the minimum knowledge preference satisfaction requirement in constraint~(\ref{P1u}c).
		However, notice that constraint~(\ref{P1u}a) or (\ref{P1u}b) may be violated during the above process, in which case we need to find the maximum-size KB in $\check{\mathcal{K}}$ by
		\begin{equation}
			n_{1}=\arg \max_{n\in \check{\mathcal{K}}}\ s_{n},
		\end{equation}
		and then reset the corresponding KBC indicators in $\bm{\alpha}_{i}^{I}$ and $\bm{\alpha}_{j}^{I}$ to $0$, i.e.,
		\begin{equation}
			\alpha_{i}^{n_{1}}=0 \quad \text{and}\quad \alpha_{j}^{n_{1}}=0.
		\end{equation}
		As a result, we obtain an initial feasible solution 
		\begin{equation}
			\bm{\alpha}_{i,j}^{I}=\left[\bm{\alpha}_{i}^{I},\bm{\alpha}_{j}^{I}\right].\label{Inisolu}
		\end{equation}
	
		\item \textit{Neighboring Solution Searching:}
		Let a $2N$-dimensional vector $\bm{\alpha}_{i,j}^{C}$ store the current solution in each search iteration, and $\mathcal{H}\left(\bm{\alpha}_{i,j}^{C}\right)$ denote its neighboring solution set, which should not include solutions that are already recorded in a tabu list, denoted as a set $\mathcal{I}$ (which will be explained later).
		Naturally, our $\mathcal{H}\left(\bm{\alpha}_{i,j}^{C}\right)$ is defined as
		\begin{equation}
			\begin{split}
    			\mathcal{H}(\bm{\alpha}_{i,j}^{C}) = \bigl\{ \bm{\alpha}_{i,j}\colon & \|\bm{\alpha}_{i,j}-\bm{\alpha}_{i,j}^{C}\| \leqslant \sigma, \\
    			&\bm{\alpha}_{i,j} \notin \mathcal{I}, \bm{\alpha}_{i,j} \in \psi \bigr\},\label{neighborhood}
			\end{split}
		\end{equation}
		where $\sigma$ is the size of the maximum neighborhood space, and $\psi$ represents the feasible solution space of $\mathbf{P1}_{i,j}$.
		Based on the above definitions, we are able to find the best solution within the current $\mathcal{H}\left(\bm{\alpha}_{i,j}^{C}\right)$ that can yield the minimum value of $\omega_{i,j}$ in an iterative fashion.
		\item \textit{Tabu List Update:}
		The tabu list $\mathcal{I}$ is a special memory mechanism for preventing subsequent searches from looping back to previously visited solutions so as to avoid trapping into the local optimum~\cite{glover1989tabu}.
		In this way, whenever there is a newly obtained $\bm{\alpha}_{i,j}^{C}$, it should be added into $\mathcal{I}$ (cannot exceed its given maximum length~\cite{salhi2002defining}).
		Besides, let $\bm{\alpha}_{i,j}^{*}$ denote another vector dedicated to storing the best solution obtained so far.
		Particularly, once $\bm{\alpha}_{i,j}^{C}$ is found to be better than $\bm{\alpha}_{i,j}^{*}$ in any iteration, it will not be added into $\mathcal{I}$ but we will have
		\begin{equation}
			\bm{\alpha}_{i,j}^{*}=\bm{\alpha}_{i,j}^{C}.
		\end{equation}
		\item \textit{Algorithm Termination Check:}
		Before commencing a new iteration, an algorithm termination criterion needs to be checked, which can be either a maximum iteration restriction or a performance improvement threshold of $\bm{\alpha}_{i,j}^{*}$ under a certain number of consecutive iterations.
	\end{itemize}
	
	\subsection{Optimal Solution to $\mathbf{P2}$}
	After the KBC sub-policy $\bm{\alpha}_{i,j}^{*}$ is found for each potential VUE $i,j$ pair, we can determine the optimal coefficient matrix $\bm{\Omega}$ to solve the VSP subproblem $\mathbf{P2}$.
	Since its objective function and two equality constraints~(\ref{P0}c) and (\ref{P0}d) are all linear, the only challenge is the $0$-$1$ constraint in~(\ref{P0}g).
	
	With regards to this, we first relax $\bm{\beta}$ into a continuous variable between $0$ and $1$ to make $\mathbf{P2}$ a linear programming problem, which can be efficiently solved with toolboxes such as CVXPY~\cite{diamond2016cvxpy}.
	Then the obtained continuous solution, denoted as $\bm{\beta}^{R}$, needs to be restored to the binary state under the original constraints.
	Here, we heuristically finalize the optimal VSP strategy $\bm{\beta}^{*}$ by
	\begin{equation}
	\label{VSPsolution}
		\beta_{i^{'}\looparrowright j^{'}}^{*}=\beta_{j^{'}\looparrowright i^{'}}^{*}=1,
	\end{equation}
	if and only if
	\begin{equation}
		\label{VSPsolution1}
		\left(i^{'},j^{'}\right)=\arg \max_{i\in \mathcal{V},j\in \mathcal{V}_{i},j>i}\beta_{i\looparrowright j}^{R}.
	\end{equation}
	Meanwhile, for the remaining $\beta_{i\looparrowright j}^{*}$ with respect to VUE $i^{'}$ and VUE $j^{'}$, we naturally have
	\begin{equation}
		\label{VSPsolution2}
			\left\{
			\begin{aligned}
			\beta_{i^{'}\looparrowright j}^{*}=\beta_{j\looparrowright i^{'}}^{*}=0, \quad & \forall j \in \mathcal{V}_{i^{'}},j \neq j^{'}\\
			\beta_{j^{'}\looparrowright i}^{*}=\beta_{i\looparrowright j^{'}}^{*}=0, \quad & \forall i \in \mathcal{V}_{j^{'}},i \neq i^{'}
			\end{aligned}.
		\right.
	\end{equation}
	Then we let $\mathcal{V}=\mathcal{V}\backslash \{i,j\}$, and repeat the above progresses until determining the optimal VSP solution for all VUEs.
	It can be seen that the number of variables is actually only $\left(\sum_{i \in \mathcal{V}}\left|\mathcal{V}_{i}\right|\right)/2$ when solving $\mathbf{P2}$, which is a fairly acceptable problem scale in practice.
	Hence, the performance compromise of our proposed heuristic VSP solution is believed to be small.
	
	\subsection{Workflow of S$^{\text{4}}$ and Complexity Analysis}
	In order to further demonstrate the full picture of the proposed solution S$^{\text{4}}$, we summarize the relevant technical points and present them in the following Algorithm 1.
		\begin{breakablealgorithm}
				\caption{The Proposed Solution S$^{\text{4}}$}
				\label{Algo1}
				\begin{algorithmic}[1]
					\REQUIRE The SCVN parameters $s_{n}$, $C_{i}$, $r_{i}^{n}$, $\xi_{i}$, $\lambda_{i}$, $\mu_{i}^{n}$, $\eta_{0}$, $\theta_{0}$
					\ENSURE The optimal KBC policy $\bm{\alpha}^{*}$ and the optimal VSP strategy $\bm{\beta}^{*}$ for each VUE $i, \forall i \in \mathcal{V}$
					\STATE Initialize $t=0$, $\tau_{i}(0)$ and $\nu(0)$ to proper positive values;
					\STATE Set the maximum number of iterations $M$ for $\mathbf{D0}$;
					\WHILE{$t<M$}
						\FOR{$i =1$ to $V$}
							\FOR{each $j \in \mathcal{V}_{i}$}
								\IF{$j>i$}
									\STATE Determine initial solution $\bm{\alpha}_{i,j}^{I}$ by (\ref{premax})-(\ref{Inisolu});
									\STATE Initialize the TS iteration as $\tilde{t}=0$, the Tabu list $\mathcal{I}(0)=\emptyset$, and $\bm{\alpha}_{i,j}^{C}(0)=\bm{\alpha}_{i,j}^{*}(0)=\bm{\alpha}_{i,j}^{I}$;
									\STATE Set the neighborhood size $\sigma$ and the maximum number of iterations $Z$ for solving each $\mathbf{P1}_{i,j}$;
									\WHILE{$\tilde{t}<Z$}
										\STATE Determine $\mathcal{H}\left(\bm{\alpha}_{i,j}^{C}\left(\tilde{t}\right)\right)$ by (\ref{neighborhood});
										\STATE Find the best feasible solution in $\mathcal{H}\left(\bm{\alpha}_{i,j}^{C}\left(\tilde{t}\right)\right)$ and assign it to $\bm{\alpha}_{i,j}^{C}(\tilde{t}+1)$;
										\IF{$\bm{\alpha}_{i,j}^{C}(\tilde{t}+1)$ is better than $\bm{\alpha}_{i,j}^{*}(\tilde{t})$}
											\STATE Update $\bm{\alpha}_{i,j}^{*}(\tilde{t}+1)=\bm{\alpha}_{i,j}^{C}(\tilde{t}+1)$;
											\STATE Keep $\mathcal{I}\left(\tilde{t}+1\right)=\mathcal{I}\left(\tilde{t}\right)$;
										\ELSE
											\STATE Keep $\bm{\alpha}_{i,j}^{*}(\tilde{t}+1)=\bm{\alpha}_{i,j}^{*}(\tilde{t})$;
											\STATE Update $\mathcal{I}\left(\tilde{t}+1\right)=\mathcal{I}\left(\tilde{t}\right)\cup\left\{\bm{\alpha}_{i,j}^{C}(\tilde{t}+1)\right\}$;
										\ENDIF
										\STATE Update $\tilde{t}=\tilde{t}+1$;
									\ENDWHILE
									\STATE Compute $\omega_{i,j}^{*}$ by substituting $\bm{\alpha}_{i,j}^{*}(Z)$ into~(\ref{VUEpaircost});
									\STATE Assign $\omega_{j,i}^{*}=\omega_{i,j}^{*}$;
								\ENDIF
							\ENDFOR
						\ENDFOR
						\STATE Renew the optimal coefficient matrix $\bm{\Omega}\left(t\right)$ by~(\ref{optimatrix});
						\STATE Solve the relaxed $\mathbf{P2}$ by CVXPY and obtain $\bm{\beta}^{R}\left(t\right)$;
						\STATE Finalize $\bm{\beta}^{*}\left(t\right)$ by~(\ref{VSPsolution})-(\ref{VSPsolution2});
						\STATE Finalize $\bm{\alpha}^{*}\left(t\right)$ by feeding $\bm{\beta}^{*}\left(t\right)$ back into $\bm{\Omega}\left(t\right)$;
						\STATE Update $\tau_{i}(t+1)$ by~(\ref{lagupdate});
						\STATE Update $\nu(t+1)$ under a given rule;
						\STATE Update $t=t+1$;
					\ENDWHILE
			\end{algorithmic}
		\end{breakablealgorithm}

	In line with this algorithm, the main flow of S$^{\text{4}}$ working in the practical SCVN is demonstrated as follows:
	\begin{itemize}
		\item \textit{Network Initialization:}
		In the initial phase, each VUE $i$ ($\forall i \in \mathcal{V}$) generates a Lagrange multiplier parameter $\tau_{i}$ and records all KBC-related status information, including its available KB storage (i.e., $C_{i}$), preferences for different KBs (i.e., $r_{i}^{n}, \forall n \in \mathcal{K},$ and $\xi_{i}$), and average local arrival rate as well as interpretation time for semantic data packets (i.e., $\lambda_{i}$ and $1/\mu_{i}^{n}$).
		Then, all VUEs need to upload the above parameters to the RSU for subsequent implementation.
		
		\item \textit{Optimal Policy Determination for KBC and VSP:}
		In this phase, the RSU first measures the SINR between VUEs to determine each VUE $i$'s communication neighbors, i.e., $\mathcal{V}_{i}$.
		Having these, the RSU is capable of computing the current optimal KBC sub-policy for each individual VUE $i,j$ pair ($j \in \mathcal{V}_{i}$) (referring to steps 4-29 in Algorithm 1).
		Afterward, the current optimal VSP and KBC policies can be jointly obtained in steps 30-33.
		Nevertheless, the Lagrange multiplier of each VUE is required to be updated with a given rule (steps 34-36), thus the RSU should repeat the procedures in steps 4-33 until satisfying the algorithm termination criterion, so as to finalize the optimal KBC and VSP policies $\bm{\alpha}^{*}$ and $\bm{\beta}^{*}$.
		\item \textit{SemCom-Empowered Service Provisioning:}
		Once each VUE receives the feedback information, it can request and download KBs from the RSU (according to $\bm{\alpha}^{*}$), and then pair with one of its neighbors (according to $\bm{\beta}^{*}$) to provide corresponding services for each other.
	\end{itemize}
	
	Herein, it is worth pointing out that the above workflow of S$^{\text{4}}$ is executed in a periodic timeline, and all decisions to update relevant parameters should be made at the end of each period. Besides, it can be observed that there are only four rounds of signaling interactions to implement a completed and successive KBC and VSP process, including vehicular information collection, optimal KBC and VSP policy assignment, KB downloading, and vehicle pairing.
	For each signaling interplay, only a few bits are needed to complete the functional confirmation work, hence the overall signaling overhead should be an apparently tolerable level in practice.
	
	In terms of the computational complexity of S$^{\text{4}}$, it is first seen that for a single $\mathbf{P1}_{i,j}$, each feasible KBC solution within $\mathcal{H}\left(\bm{\alpha}_{i,j}^{C}\right)$ needs to be computed once in any of its iterations.
	Combining that $\sigma$ is given as a small parameter compared to $N$ in~(\ref{neighborhood}), the complexity in each iteration can be estimated by $\mathcal{O}\left(\binom{2N}{\sigma}\right)=\mathcal{O}\left(N^{\sigma}\right)$.
	If the maximum number of its iterations is assumed to be $Z$, then solving each $\mathbf{P1}_{i,j}$ would require complexity $\mathcal{O}\left(ZN^{\sigma}\right)$.
	In addition, the linear programming method is utilized for the relaxed $\mathbf{P2}$, where $\mathcal{O}\left(V^{4}\right)$ complexity is needed~\cite{lee2019solving} to solve a group of $\left(\sum_{i \in \mathcal{V}}\left|\mathcal{V}_{i}\right|\right)$ VSP variables.
	Moreover, note that in any iteration of $\mathbf{D0}$, a total of $\left(\left(\sum_{i \in \mathcal{V}}\left|\mathcal{V}_{i}\right|\right)/2\right)$ subproblems $\mathbf{P1}_{i,j}$ and one subproblem $\mathbf{P2}$ need to be solved simultaneously, thereby its corresponding complexity is $\mathcal{O}\left(V^{2}ZN^{\sigma}+V^{4}\right)$.
	If denoting the maximum number of iterations that can make $\mathbf{D0}$ to converge as $M$, the proposed S$^{\text{4}}$ would have a polynomial-time overall complexity, given as $\mathcal{O}\left(MV^{2}\left(ZN^{\sigma}+V^{2}\right)\right)$.
	
	\section{Numerical Results and Discussions}
	\begin{table}[t]
		\centering
		\caption{Simulation Parameters}
		\label{SimuPara}
		\setlength{\tabcolsep}{3pt}
		\renewcommand\arraystretch{1.5}
		\begin{tabular}{|m{4.15cm}<{\raggedright}|m{4.15cm}<{\raggedright}|}\hline
			\textbf{Parameters} & \textbf{Values} \\ \hline
			Number of VUEs ($V$) & $60$ \\ \hline
			Number of KBs ($N$) & $12$ \\ \hline
			Size of KB $n$ ($s_{n}$) & $1\sim 5$ units (randomly)\\ \hline
			KB storage capacity of VUEs ($C_{i}$) & $24$ units \\ \hline
			Skewness of the Zipf distribution with respect to each VUE's KB preference ($\xi_{i}$) & $1.0$ \\ \hline
			Average arrival rate of total semantic data packets of VUEs ($\lambda_{i}$) & $100$ packets/s~\cite{7434039} \\ \hline
			Average interpretation time of KB $n$-based semantic data packets of VUEs ($1/\mu_{i}^{n}$) & $5\times 10^{-3}\sim 1\times 10^{-2}$ s/packet (randomly) \\ \hline
			Cell radius of the RSU & $500$ m \\ \hline
			VUE drop model & Spatial Poisson process~\cite{3GPPLTE} \\ \hline
			Number of lanes & $3$ in each direction ($6$ in total) \\ \hline
			Lane width & $4$ m \\ \hline
			Absolute velocity of VUEs & $70$ km/h~\cite{liang2017resource} \\ \hline
			Density of VUEs & Average inter-vehicle distance is $2.5$ sec $\times$ absolute velocity of VUEs~\cite{3GPPLTE} \\ \hline
			Transmit power of VUEs & $20$ dBm~\cite{zhang2021uav} \\ \hline
			Noise power & $-114$ dBm \\ \hline
			Path loss model & $128.1+37.6\log\left(d\ \text{[km]}\right)$~\cite{ding2022two} \\ \hline
			Channel fading model & Log-normal shadowing distribution with standard deviation of $8$ dB $+$ Rayleigh fast fading~\cite{ding2022two} \\ \hline
			Minimum knowledge preference satisfaction threshold ($\eta_{0}$) & $0.5$ \\ \hline
			Maximum knowledge mismatch degree threshold ($\theta_{0}$) & $0.1$ \\ \hline
		\end{tabular}
	\end{table}
	In this section, numerical evaluations are conducted to demonstrate the performance of the proposed solution S$^{\text{4}}$ in SCVNs, where we employ Python 3.7-based PyCharm as the simulator platform and implement it in a computer with six CPU cores and Inter Core i7 processor.
	To preserve generality, we model a multi-lane freeway passing through a single cell with the RSU at its center, and multiple VUEs are dropped on the lanes according to the spatial Poisson process~\cite{3GPPLTE}.
	The main parameters of the basic network setup as well as their corresponding values can be found in Table~\ref{SimuPara}~\cite{3GPPLTE,7434039,liang2017resource,zhang2021uav,ding2022two}.
	As for the settings relevant to SemCom, a total of $12$ different KBs are preset to provide VUEs with a variety of distinct services, and each of them has a storage size randomly distributed from $1$ to $5$ units.
	Correspondingly, we set a uniform KB storage capacity of $24$ units for all VUEs.
	Besides, each VUE's preference ranking for all KBs (i.e., $r_{i}^{n}$) is generated in an independent and random manner, where their respective Zipf distributions are assumed to have the same skewness $1.0$.
	Likewise, either the average arrival rate of total semantic data packets, or the average interpretation time for packets based on the same KB $n$, is considered to be the same for all VUEs.
	Here, we fix the average total arrival rate $\lambda_{i}$ at $100$ packets/s and randomly generate the value of $1/\mu_{i}^{n}$ in a range of $5\times 10^{-3}\sim 1\times 10^{-2}$ s/packet with respect to different KB $n$, as the average interpretation time of different KBs-based packets is different from each other, which is to guarantee the steady-state of the queuing system at each VUE pair, as has mentioned in Footnote 4.
	Further, the minimum knowledge preference satisfaction threshold $\eta_{0}$ and the maximum knowledge mismatch degree threshold $\theta_{0}$ are prescribed as $0.5$ and $0.1$, respectively.
	Notably, all these parameter values in Table~\ref{SimuPara} are set by default unless otherwise specified, and all subsequent numerical results are obtained by averaging over a sufficiently large number of trials.
	
	For comparison purposes, we utilize two different benchmark schemes of SemCom-empowered service provisioning herein: 1) Distance-first pairing (DFP) strategy which assumes each VUE to choose its nearest unpaired VUE for V2V pairing; 2) Knowledge-first pairing (KFP) in which each VUE selects its neighboring unpaired VUE with the highest KB matching degree for V2V pairing.
	In the meantime, a personal preference-first KBC policy is considered for both benchmarks, which allows each VUE to construct KBs with the highest preferences until $\eta_{0}$ is satisfied, and then randomly select these unconstructed KBs until reaching respective maximum capacity.
	
	\begin{figure}[t]
		\centering
		\includegraphics[width=0.48\textwidth]{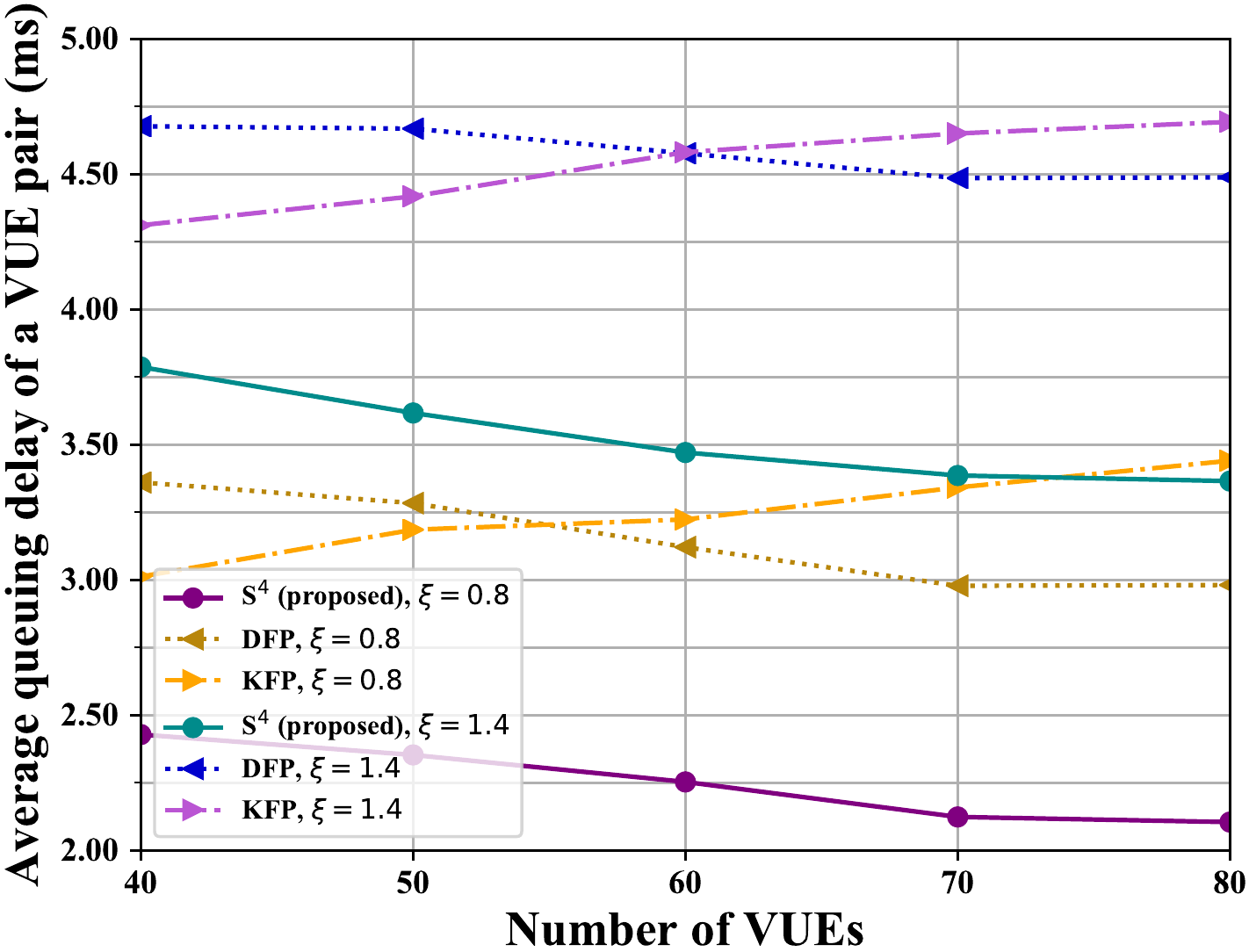}
		\caption{Average queuing latency of a VUE pair vs. varying numbers of VUEs.}
		\label{SimFig1}
	\end{figure}
    Fig.~\ref{SimFig1} first depicts the average queuing latency performance of a VUE pair against varying numbers of VUEs, where two different KB preference skewness $\xi=0.8$ and $\xi=1.4$ are considered.
    In this figure, the latency of S$^{\text{4}}$ declines at the beginning with the number of VUEs, then remains stable beyond $70$ VUEs, and it can always outperform both benchmarks with an average latency reduction of around $1$ ms at any $\xi$. 
    The rationale behind this trend is that the more neighbors each VUE can have, the better chance of achieving the low queuing latency for each VUE pair, which will be eventually stabilized when reaching the respective best achievable latency with a fixed bandwidth budget.
    Moreover, it is observed that a larger $\xi$ causes a higher latency penalty, since the vast majority of VUEs' KBC is concentrated on a small number of KBs when $\xi$ increases.
    Clearly, a larger $\xi$ will make each participant more difficult to find the best VUE with the low latency under the given knowledge mismatching requirement $\theta_{0}$, thus resulting in a degraded performance.
    
	\begin{figure}[t]
		\centering
		\includegraphics[width=0.48\textwidth]{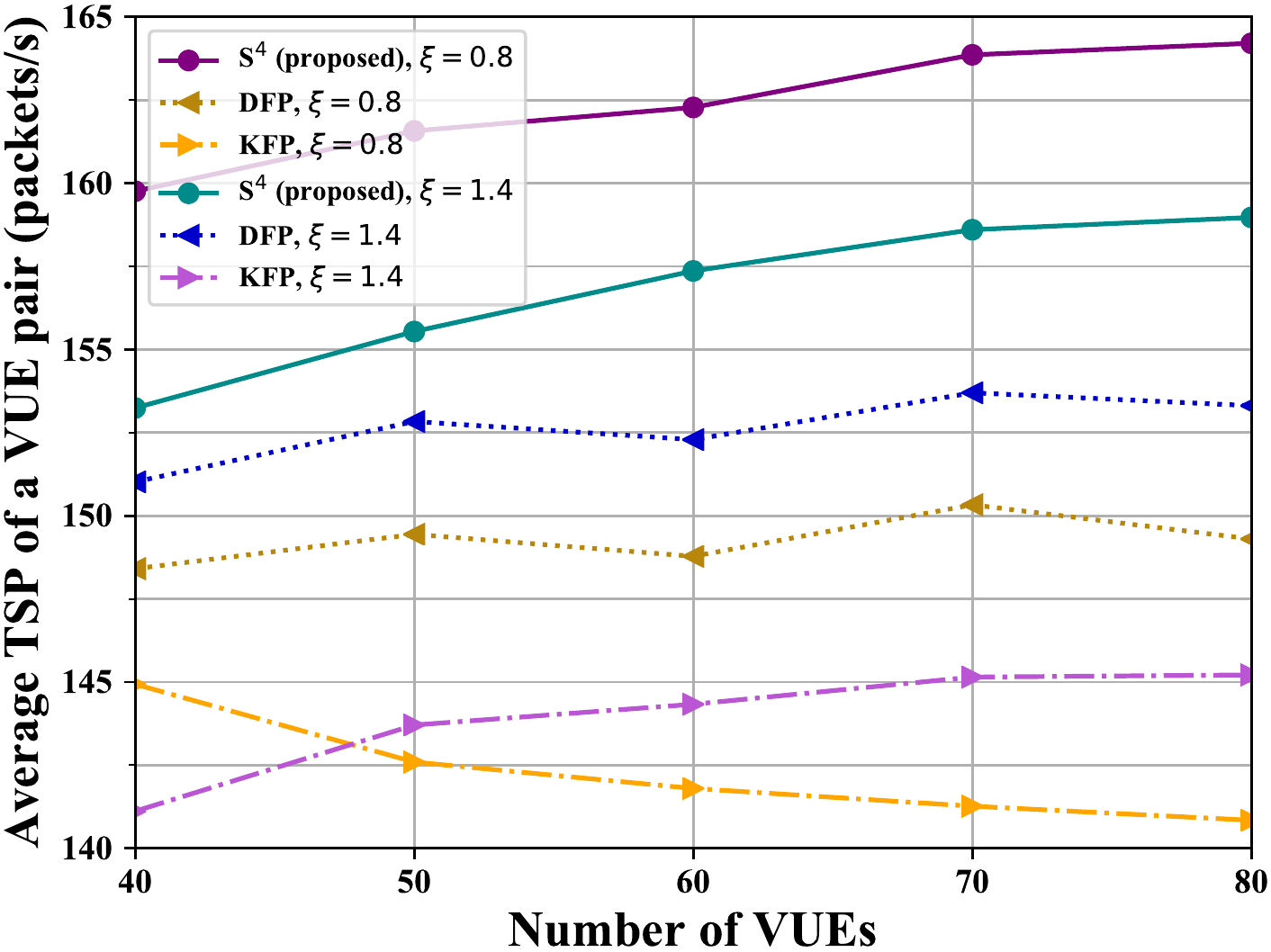}
		\caption{Average TSP of a VUE pair vs. varying numbers of VUEs.}
		\label{SimFig2}
	\end{figure}
    The above analysis also applies to Fig.~\ref{SimFig2}, which compares all the three methods under the same settings as Fig.~\ref{SimFig1} to demonstrate the performance of the average throughput in semantic packets (TSP).
    Specifically, the TSP represents the total number of semantic packets that can be interpreted by a VUE pair per second, whose value is determined based on $\mathds{E}\left[W_{i\looparrowright j}\right]$ in~(\ref{PK1}).
    Likewise, a higher TSP is obtained as the number of VUEs increases, and our S$^{\text{4}}$ is still far better than the two benchmarks at any point, e.g., with an average performance gain of $14$ packets/s compared with DFP and $20$ packets/s with KFP at $\xi=0.8$.
    Again, we see a better TSP when the KB popularity is diluted by a smaller $\xi$.
    
    \begin{figure}[t]
		\centering
		\includegraphics[width=0.48\textwidth]{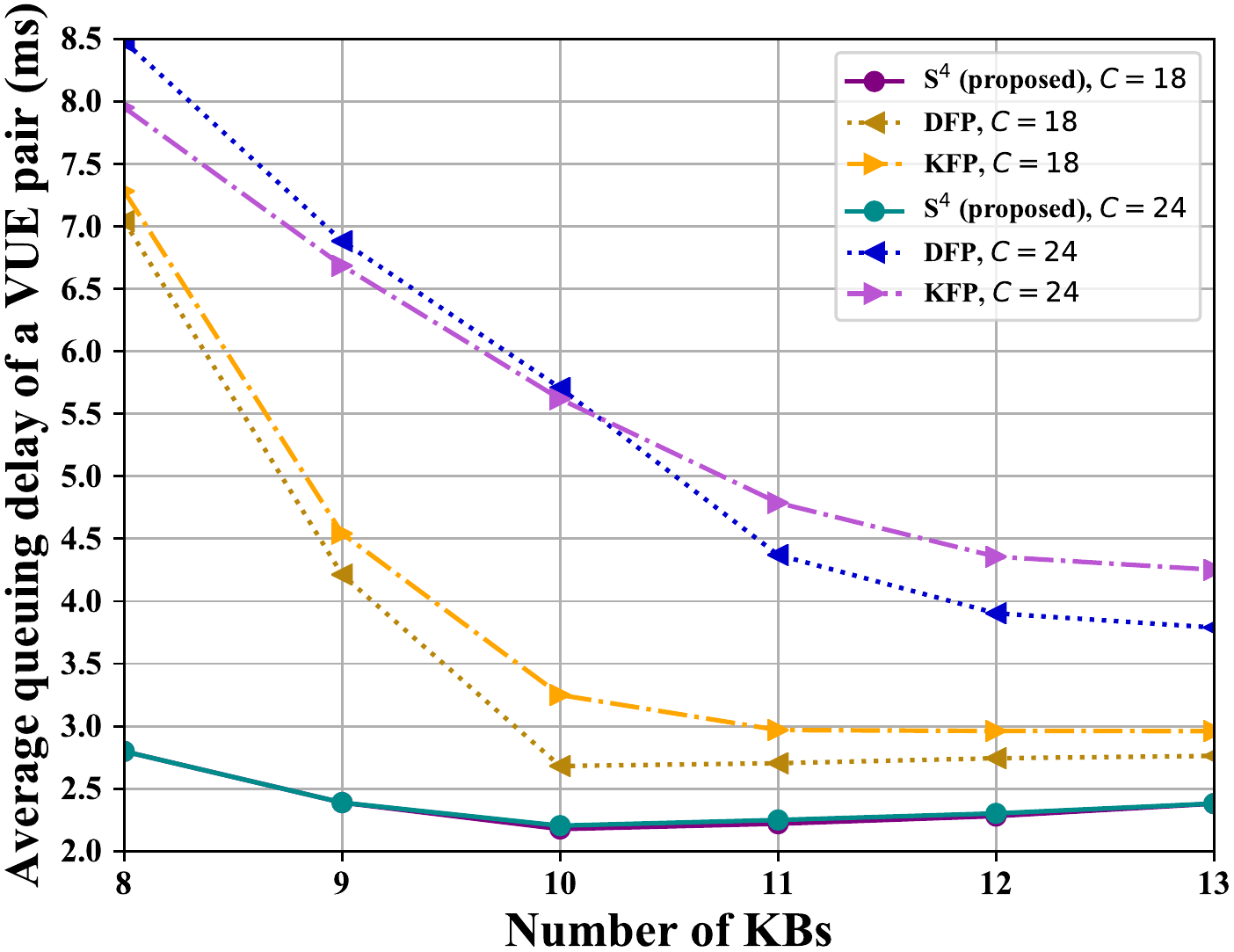} 
		\caption{Average queuing latency of a VUE pair with varying numbers of KBs.}
		\label{SimFig3}
    \end{figure}
    
    \begin{figure}[t]
		\centering
		\includegraphics[width=0.48\textwidth]{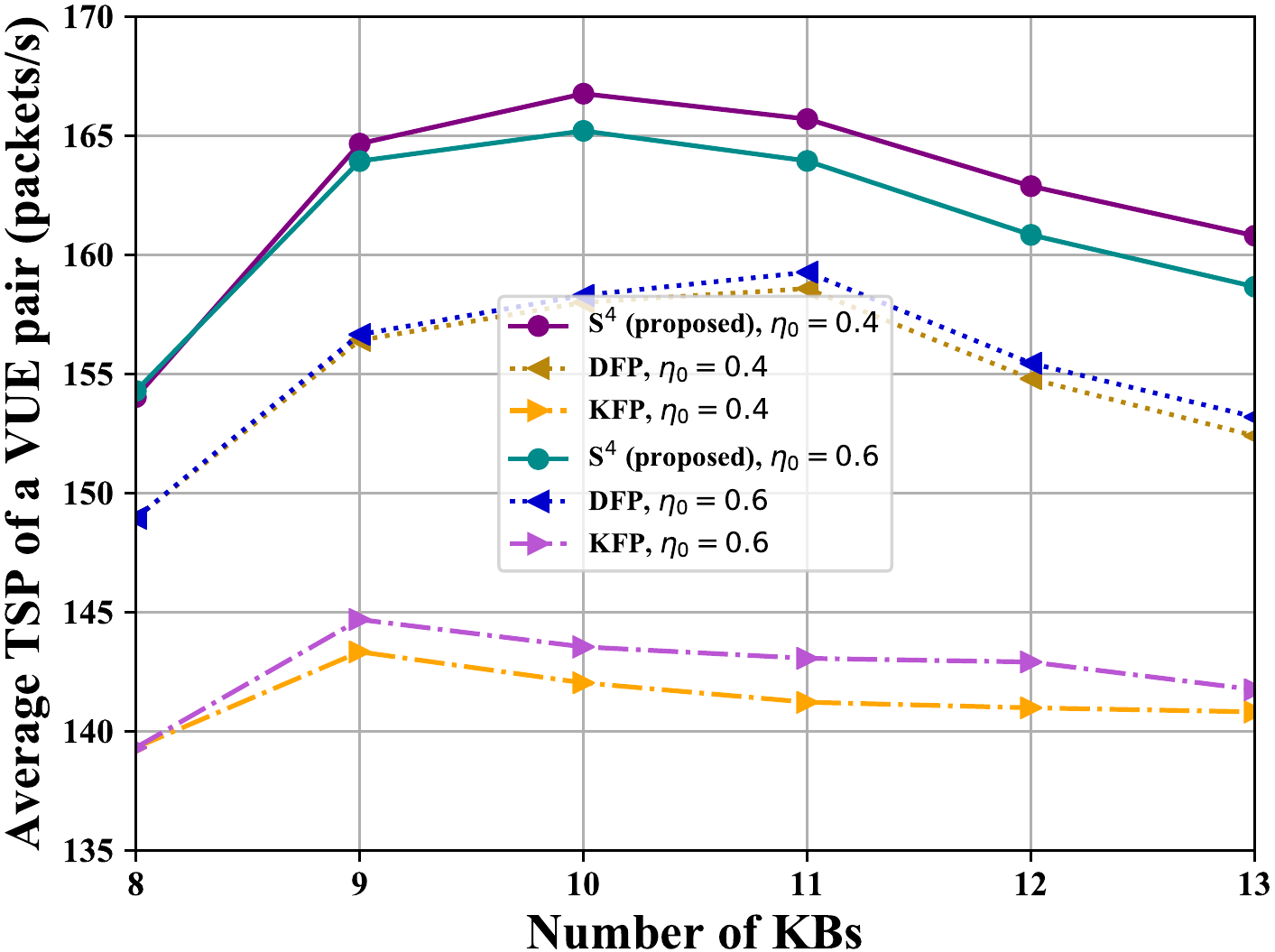} 
		\caption{Average TSP of a VUE pair with varying numbers of KBs.}
		\label{SimFig4}
    \end{figure}
    
    Next, we explore the impact of varying number of KBs on the average queuing latency of a VUE pair with different VUE capacities $C=18$ and $C=24$, as demonstrated in Fig.~\ref{SimFig3}.
    It can be found that the latency drops fast at the beginning, and then rises slightly after exceeding $10$ KBs, whereas the performance of our S$^{\text{4}}$ still surpasses the benchmarks.
    This trend is attributed to the fact that more KBs imply less discrepancy in VUEs' preferences for different KBs given the fixed $\xi$, thereby at first leading to the higher probability for two paired VUEs constructing the KBs with high interpretation rates so as to render a lower delay.
    However, such performance gains will be saturated and even worsen when these KBs with low interpretation rates become inevitably dominant in order to meet the minimum knowledge preference satisfaction threshold $\eta_{0}$.
    Besides, it is seen that different VUE capacities have little effect on the latency of S$^{\text{4}}$, although the larger capacity can construct more KBs.
	This is due to the latency-minimization objective we particularly focus on in the delay-sensitive SCVN, and only the KBs with low interpretation time should be selected.
    Afterward, we draw the TSP performance against different numbers of KBs with $\eta_{0}=0.4$ and $\eta_{0}=0.6$, as shown in Fig.~\ref{SimFig4}.
    The similar trend to Fig.~\ref{SimFig3} is observed here as well, i.e., the TSP of S$^{\text{4}}$ rises at the beginning and then falls after $10$ KBs, and a much higher TSP is provided compared with the benchmarks.
    Meanwhile, it is noticed that a lower $\eta_{0}$ brings a better TSP, as fewer KBs need to be constructed to guarantee the high average interpretation rates in the queue, as mentioned earlier.
    
    \begin{figure}[t]
		\centering
		\includegraphics[width=0.48\textwidth]{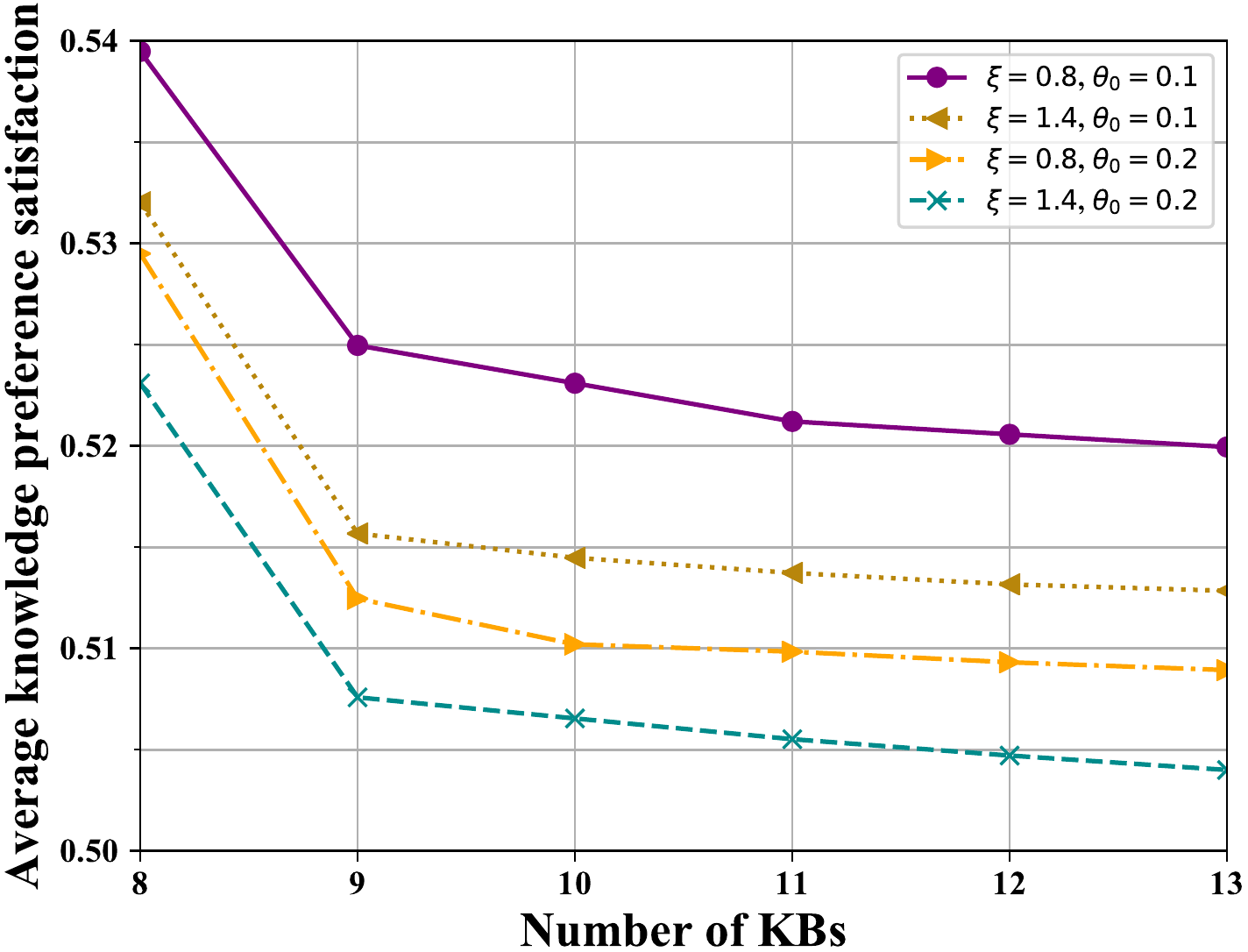} 
		\caption{Average knowledge preference satisfaction reached at a VUE with varying numbers of KBs.}
		\label{SimFig5}
    \end{figure}
    In addition, we validate the average knowledge preference satisfaction $\bar{\eta}=\frac{1}{V}\sum_{i \in \mathcal{V}}\eta_{i}$ reached at each VUE with varying numbers of KBs as shown in Fig.~\ref{SimFig5}, where $\xi = 0.8$, $\xi = 1.4$, $\theta_{0}=0.1$, and $\theta_{0}=0.2$ are taken into account.
    As the number of KBs increases, a lower $\bar{\eta}$ is obtained, which is to prevent these unnecessary KBs from being constructed while satisfying $\eta_{0}$ to the greatest extent.
    For the two curves with different $\xi$, referring to the analysis of Fig.~\ref{SimFig1}, a higher $\xi$ indicates a more concentrated KB preference, which means some extra KBs need to be constructed to meet the maximum $\theta_{0}$ requirement.
   	 Because of this, we also see a lower $\bar{\eta}$ at a higher $\theta_{0}$, since a more tolerable knowledge mismatch degree is more likely to avoid the unnecessary KBC.
   	
     \begin{figure}[t]
		\centering
		\includegraphics[width=0.48\textwidth]{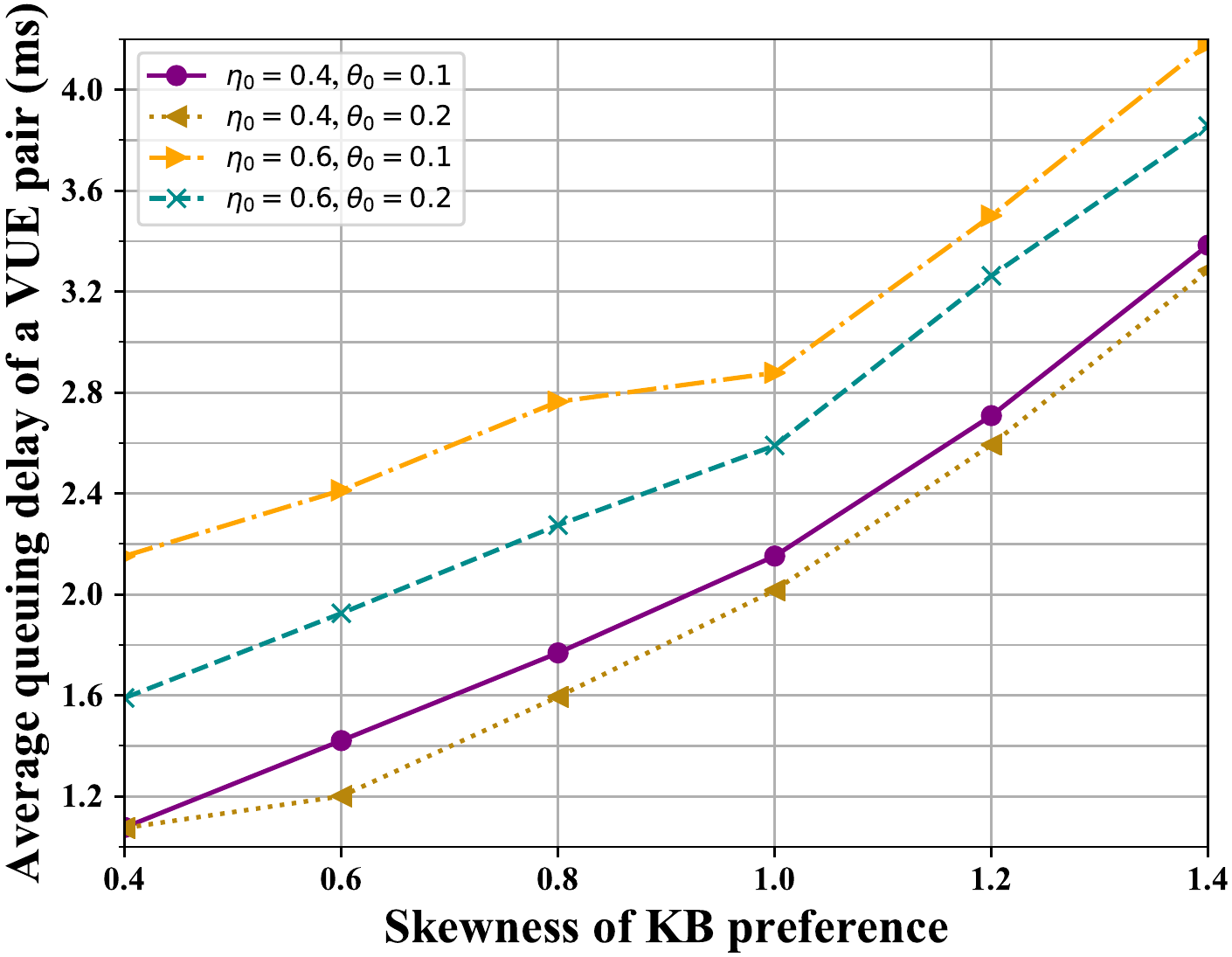} 
		\caption{Average queuing latency of a VUE pair with varying skewness of VUEs' KB preferences.}
		\label{SimFig6}
    \end{figure}
    
    \begin{figure}[t]
		\centering
		\includegraphics[width=0.48\textwidth]{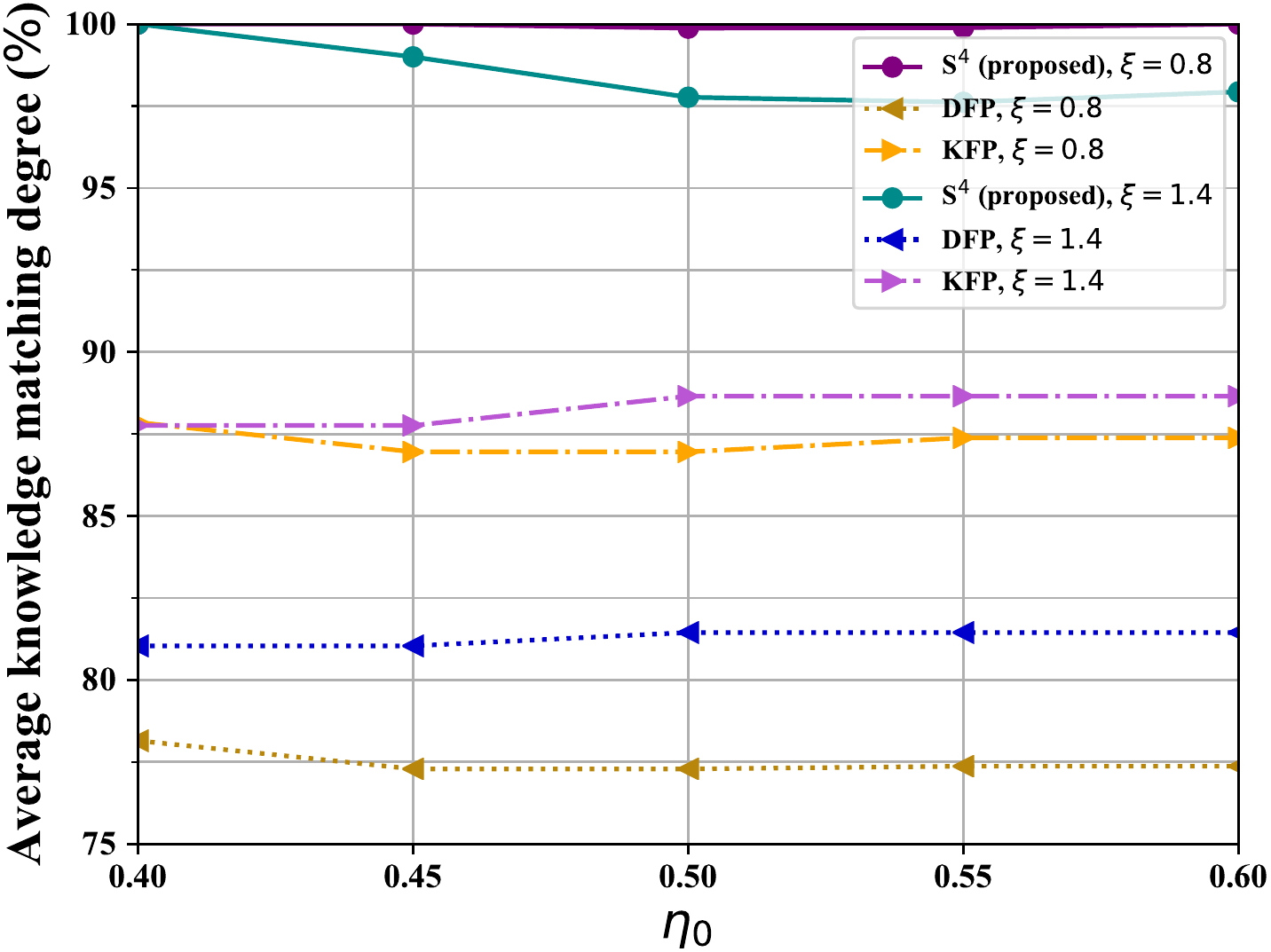} 
		\caption{Average knowledge matching degree of a VUE pair vs. different knowledge preference satisfaction requirements.}
		\label{SimFig7}
    \end{figure}	
	Fig.~\ref{SimFig6} presents the effect of varying $\xi$ on the average queuing delay compared between different $\eta_{0}$ and $\theta_{0}$.
	As expected, the latency of all three methods increases with $\xi$, and such an upward trend is consistent with the previous results.
	Specially, it is observed that the curve of $\eta_{0}=0.4$ and $\theta_{0}=0.2$ brings the best latency performance.
	This can be understood as that either the lower satisfaction threshold or the higher mismatch tolerance can avoid the construction of unnecessary KBs with low interpretation rates, as discussed before.
	Naturally, the better latency performance is obtained when the constraints become less stringent.

	Finally, we plot the average knowledge matching degree, defined as $\bar{\rho}=\frac{1}{V}\sum_{i \in \mathcal{V}, j \in \mathcal{V}_{i}}\left(1-\theta_{i \looparrowright j}\right)$, with two different $\xi$ in Fig.~\ref{SimFig7}.
	It can be seen that under the threshold of $\theta_{0}=0.1$, the proposed solution S$^{\text{4}}$ is always above a $97.5$\% match degree, which is much higher than that of benchmarks.
	Furthermore, a higher $\xi$ causes the slight drop of $\bar{\rho}$, which exactly proves the conclusion of Fig.~\ref{SimFig1}, i.e., a more concentrated KB preference makes it more difficult for VUEs to find a highly matching neighbor in the VSP phase, especially for the one that can bring lower latency and meet all constraints at the same time.
	
	\section{Conclusions}
	In this paper, we proposed a novel solution S$^{\text{4}}$ to address the SemCom-empowered service provisioning problem in the SCVN.
	To align with the stringent xURLLC requirements, the KB matching based queuing latency expression of semantic data packets was first derived, and then we identified and formulated the fundamental problem of KBC and VSP to minimize the queuing latency for all VUE pairs.
	After the primal-dual problem transformation, a two-stage method was developed specifically to jointly solve multiple subproblems related to KBC and VSP with low computational complexity, and the solution optimality has been theoretically proved.
	Numerical results verified the sufficient performance superiority of S$^{\text{4}}$ in terms of both latency and reliability by comparing it with two different benchmarks.

	This work can be served as a pioneer in exploring the potential of applying SemCom in vehicular networks for provisioning pertinent communication services with the aim of meeting the xURLLC requirements.
	Besides, other advanced networking issues in the SCVN, such as semantic-aware resource allocation or semantic transceiver design, can treat this work as the fundamental theoretical framework for reference.
	Since this work is limited to determining optimal instantaneous KBC and VSP policies for VUEs with known KB popularity, the further problem in an expanded SCVN scenario of considering high user mobility and unaware user preferences will be investigated in our future research.
	
	\begin {appendices}
		\section{Proof of Proposition 1}
		Given the optimal KBC solution $\bm{\alpha}^{*}$, let $\bm{\beta}^{*}=\left[\beta_{1\looparrowright j^{*}_{1}},\beta_{2\looparrowright j^{*}_{2}},\cdots,\beta_{V\looparrowright j^{*}_{V}}\right]^{T}$ be the corresponding optimal VSP solution to the problem in~(\ref{Dual}) under the same dual variable $\bm{\tau}$, where $\beta_{i\looparrowright j^{*}_{i}}$ ($\forall i \in \mathcal{V}$) indicates that VUE $j^{*}_{i}$ is the optimal SemCom node for VUE $i$, i.e., $\beta_{i\looparrowright j^{*}_{i}}=1$.
		
		From $\omega_{i,j}$ defined in~(\ref{VUEpaircost}), the objective function $\widetilde{L}_{\bm{\tau}}(\bm{\alpha},\bm{\beta})$ in~(\ref{Dual}) can be rewritten as
		\begin{equation}
		\label{objective1}
			\begin{aligned}
				\widetilde{L}_{\bm{\tau}}(\bm{\alpha},\bm{\beta})=&\frac{1}{2}\sum_{i \in \mathcal{V}}\sum_{j \in \mathcal{V}_{i}}\beta_{i\looparrowright j}\omega_{i,j}=\sum_{i \in \mathcal{V}}\sum_{j \in \mathcal{V}_{i},j>i}\beta_{i\looparrowright j}\omega_{i,j},
			\end{aligned}
		\end{equation}
		then we substitute $\bm{\beta}^{*}$ into~(\ref{objective1}) and yield
		\begin{equation}
		\label{objective2}
			\begin{aligned}
				\widetilde{L}_{\bm{\tau}}(\bm{\alpha},\bm{\beta}^{*})=&\frac{1}{2}\sum_{i \in \mathcal{V}}\omega_{i,j^{*}_{i}}=\sum_{i \in \mathcal{V},i<j^{*}_{i}}\omega_{i,j^{*}_{i}},
			\end{aligned}
		\end{equation}
		where $\omega_{i,j^{*}_{i}}$ is the term only related to VUE $i,j^{*}_{i}$ pair.
		
		Undoubtedly, if $\bm{\alpha}^{*}$ is further substituted into~(\ref{objective2}), we can straightforwardly reach the optimality of the problem in~(\ref{Dual}).
		Since different VUE $i,j^{*}_{i}$ pairs are independent of each other, it means that different terms related to $\omega_{i,j^{*}_{i}}$ are independent of each other as well in $\widetilde{L}_{\bm{\tau}}(\bm{\alpha},\bm{\beta}^{*})$.
		Therefore, we can directly draw an important conclusion that achieving the optimality of $\widetilde{L}_{\bm{\tau}}(\bm{\alpha},\bm{\beta}^{*})$ is equivalent to achieving the optimality of each $\omega_{i,j^{*}_{i}}$, where the optimality can be reached when $\bm{\alpha}=\bm{\alpha}^{*}$.
		
		In view of the above, we know that $\bm{\alpha}^{*}_{i}$ must be the optimal solution of $\omega_{i,j^{*}_{i}},\forall i \in \mathcal{V}$.
		Further combined with another fact that $\omega_{i,j}$ is the objective of $\mathbf{P1}_{i,j},\forall \left( i,j\right) \in \mathcal{V}\times \mathcal{V}_{i}, j>i$ where $\bm{\alpha}^{*}_{i_{(j)}}$ is the corresponding optimal solution, we ensure that the equality $\bm{\alpha}^{*}_{i_{(j)}}=\bm{\alpha}^{*}_{i}$ holds when $j=j^{*}_{i}$.
		
		\section{Proof of Proposition 2}
		Suppose that $\left(\bm{\alpha}^{*},\bm{\beta}^{*}\right)$ is not optimal for the problem in~(\ref{Dual}), which means there must exist another solution, denoted as $\bar{\bm{\alpha}}=\left[\bar{\bm{\alpha}}_{1}, \bar{\bm{\alpha}}_{2}, \cdots,\bar{\bm{\alpha}}_{V}\right]^{T}$ and $\bar{\bm{\beta}}=\left[\beta_{1\looparrowright \bar{j}_{1}},\beta_{2\looparrowright \bar{j}_{2}},\cdots,\beta_{V\looparrowright \bar{j}_{V}}\right]^{T}$, such that 
		\begin{equation}
			\widetilde{L}_{\bm{\tau}}(\bar{\bm{\alpha}},\bar{\bm{\beta}})<\widetilde{L}_{\bm{\tau}}(\bm{\alpha}^{*},\bm{\beta}^{*}).\label{B1}
		\end{equation}
		
		On the one hand, since $\bm{\beta}^{*}$ is the optimal solution to $\mathbf{P2}$, for $\bar{\bm{\beta}}\neq \bm{\beta}^{*}$, we have $\widetilde{L}_{\bm{\tau}}(\bm{\alpha}^{*},\bm{\beta}^{*})<\widetilde{L}_{\bm{\tau}}(\bm{\alpha}^{*},\bar{\bm{\beta}})$.
		On the other hand, directly applying the conclusions in Proposition 1, it is seen that $\forall i \in \mathcal{V}$, we have $\bm{\alpha}^{*}_{i_{(j)}}=\bar{\bm{\alpha}}_{i}$ when $j=\bar{j}_{i}$.
		Combined with the previous assumption that $\bar{\bm{\beta}}$ is the optimal VSP solution to problem~(\ref{Dual}), $\omega^{*}_{i,\bar{j}_{i}}=\bar{\omega}_{i,\bar{j}_{i}}$ holds such that
		\begin{equation}
			\widetilde{L}_{\bm{\tau}}(\bar{\bm{\alpha}},\bar{\bm{\beta}})=\widetilde{L}_{\bm{\tau}}(\bm{\alpha}^{*},\bar{\bm{\beta}})>\widetilde{L}_{\bm{\tau}}(\bm{\alpha}^{*},\bm{\beta}^{*}).\label{B2}
		\end{equation}
		
		However, there is a contradiction between~(\ref{B1}) and~(\ref{B2}).
		Consequently, the assumption cannot hold, which means that $\left(\bm{\alpha}^{*},\bm{\beta}^{*}\right)$ is exactly the optimal solution to problem~(\ref{Dual}).
		
	\end {appendices}
	
	\section*{Acknowledgment}
	The authors would like to thank Dr. Rafiq Swash and Dr. Wallizada Mohibullah from AIDrivers Ltd. for their comments, which have substantially improved the quality of this paper.
	
	\bibliographystyle{IEEEtran}
	\bibliography{main}

	\begin{IEEEbiography}[{\includegraphics[width=1in, height=1.25in, clip, keepaspectratio]{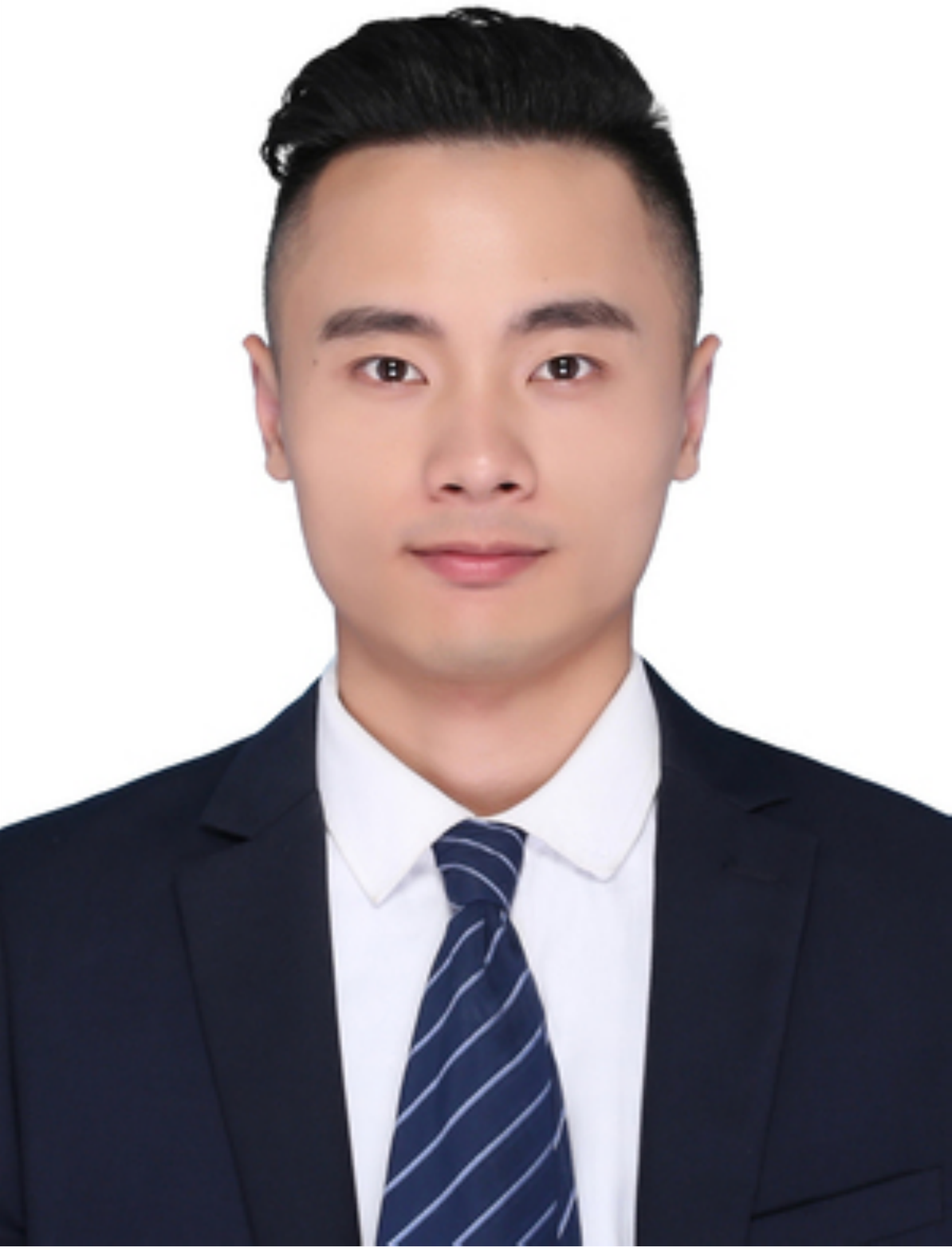}}]{Le Xia} (Graduate Student Member, IEEE)
	 received the B.Eng. degree in communication engineering and the M.Eng. degree in electronics and communication engineering from the University of Electronic Science and Technology of China (UESTC) in 2017 and 2020, respectively. He is currently pursuing his Ph.D. degree with James Watt School of Engineering, the University of Glasgow, United Kingdom. His research interests include semantic communications, intelligent vehicular networks, and resource management in next-generation wireless networks.
	\end{IEEEbiography}

	\begin{IEEEbiography}[{\includegraphics[width=1in, height=1.25in, clip, keepaspectratio]{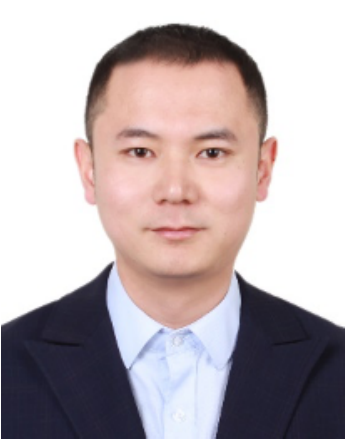}}]{Yao Sun} (Senior Member, IEEE)
	is currently a Lecturer with James Watt School of Engineering, the University of Glasgow, Glasgow, UK. Dr Sun has won the IEEE Communication Society of TAOS Best Paper Award in 2019 ICC, IEEE IoT Journal Best Paper Award 2022 and Best Paper Award in 22nd ICCT. His research interests include intelligent wireless networking, semantic communications, blockchain system, and resource management in next generation mobile networks. Dr. Sun is a senior member of IEEE.
	\end{IEEEbiography}

	\begin{IEEEbiography}[{\includegraphics[width=1in, height=1.25in, clip, keepaspectratio]{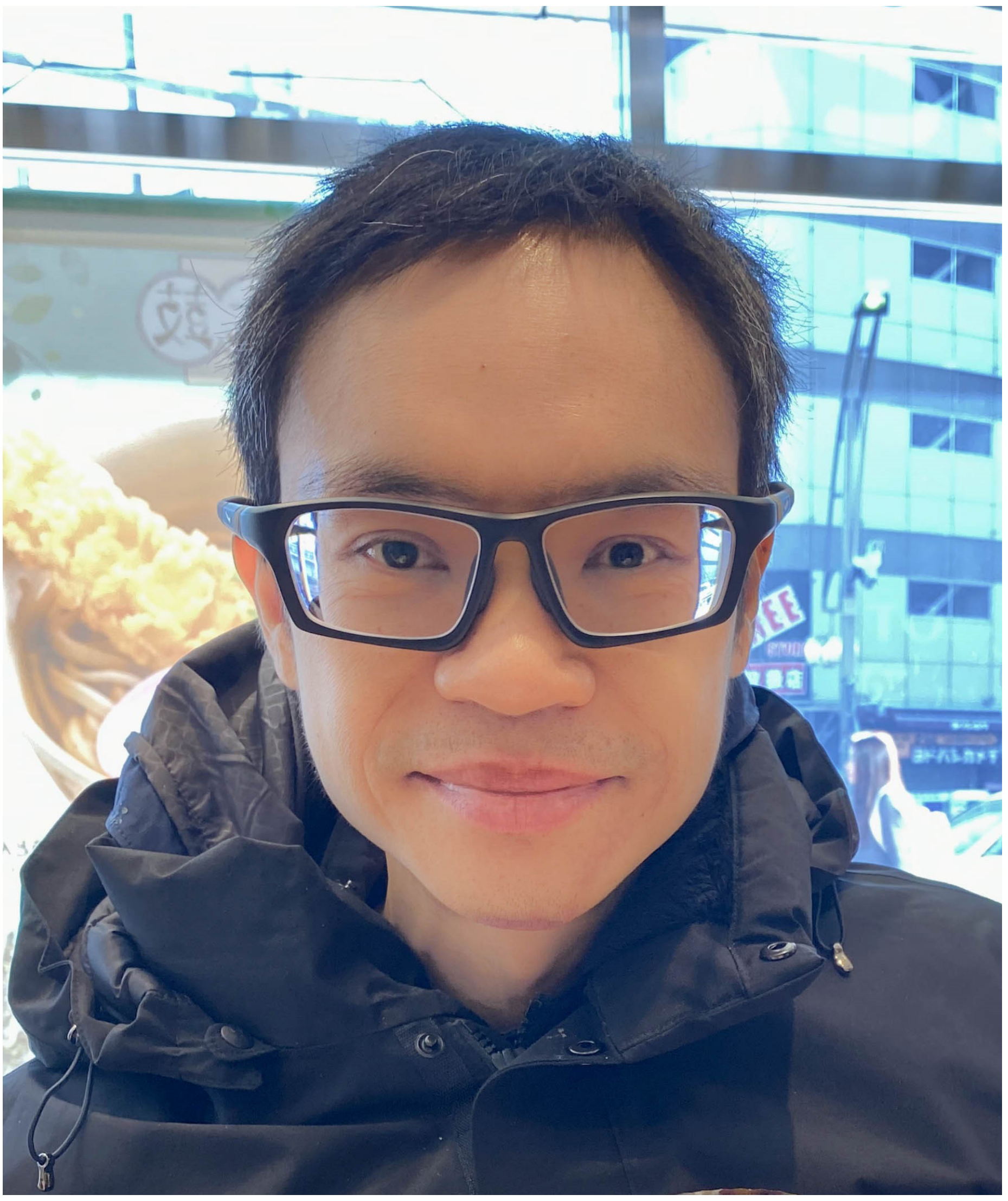}}]{Dusit Niyato} (M'09-SM'15-F'17, IEEE)
	  is a professor in the School of Computer Science and Engineering, at Nanyang Technological University, Singapore. He received B.Eng. from King Mongkut's Institute of Technology Ladkrabang (KMITL), Thailand in 1999 and Ph.D. in Electrical and Computer Engineering from the University of Manitoba, Canada in 2008. His research interests are in the areas of sustainability, edge intelligence, decentralized machine learning, and incentive mechanism design.
	\end{IEEEbiography}

	\begin{IEEEbiography}[{\includegraphics[width=1in, height=1.25in, clip, keepaspectratio]{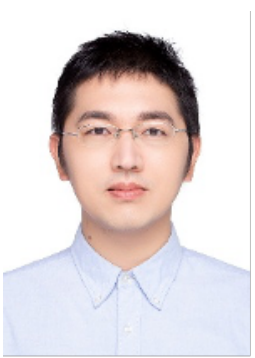}}]{Daquan Feng} (Member, IEEE)
	  is now an associate professor with the Shenzhen Key Laboratory of Digital Creative Technology, the Guangdong Province Engineering Laboratory for Digital Creative Technology, the Guangdong-Hong Kong Joint Laboratory for Big Data Imaging and Communication, College of Electronics and Information Engineering, Shenzhen University, Shenzhen, China. His research interests include URLLC communications, MEC, and massive IoT networks. Dr. Feng is an Associate Editor of IEEE Communications Letters, ICT Express, and Digital Communications and Networks.
	\end{IEEEbiography}
	
	\begin{IEEEbiography}[{\includegraphics[width=1in, height=1.25in, clip, keepaspectratio]{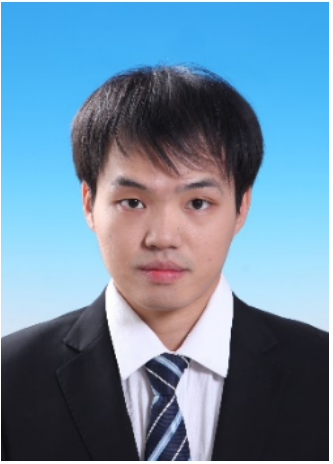}}]{Lei Feng} (Member, IEEE)
	  received his B.Eng. and Ph.D. degrees in Communication and Information Systems from the Beijing University of Posts and Telecommunications (BUPT) in 2009 and 2015, respectively. He is currently an Associate Professor at present in State Key Laboratory of Networking and Switching Technology, BUPT. His research interests include knowledge-driven next generation networks, 5G time-sensitive network technologies for Industrial IoT, and resource management and coordination of mobile networks.
	\end{IEEEbiography}
	\vspace{-430pt}
	\begin{IEEEbiography}[{\includegraphics[width=1in, height=1.25in, clip, keepaspectratio]{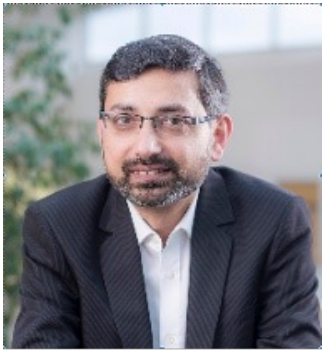}}]{Muhammad Ali Imran} (Fellow, IEEE) 
	received his M.Sc. (Distinction) and Ph.D. degrees from Imperial College London, UK, in 2002 and 2007, respectively. He is a Professor in Communication Systems in the University of Glasgow. He has a global collaborative research network spanning both academia and key industrial players in the field of wireless communications. He has supervised 50+ successful PhD graduates and published over 600 peer-reviewed research papers including more than 100 IEEE Transaction papers. Prof. Imran is a Fellow of IEEE.
	\end{IEEEbiography}
\end{document}